\documentclass[lettersize,journal]{IEEEtran}

\usepackage{amsmath,amsfonts}
\usepackage{algorithmic}
\usepackage{array}
\usepackage[caption=false,font=normalsize,labelfont=sf,textfont=sf]{subfig}
\usepackage{textcomp}
\usepackage{stfloats}
\usepackage{url}
\usepackage{verbatim}
\usepackage{graphicx}
\def\BibTeX{{\rm B\kern-.05em{\sc i\kern-.025em b}\kern-.08em
    T\kern-.1667em\lower.7ex\hbox{E}\kern-.125emX}}
\usepackage{balance}

\usepackage{enumitem}

\usepackage{multirow}
\usepackage{tabularx}

\usepackage{color}
\usepackage[table]{xcolor}
\usepackage{xcolor}

\usepackage{amsmath,amssymb,amsfonts}
\usepackage{bm}
\usepackage{bbding}

\usepackage{hyperref}

\begin{document}

\title{Enabling AutoML for Zero-Touch Network Security: Use-Case Driven Analysis}

\author{Li Yang, \IEEEmembership{Member, IEEE}, Mirna El Rajab, \IEEEmembership{Student Member, IEEE}, Abdallah Shami, \IEEEmembership{Senior Member, IEEE},\\ and Sami Muhaidat, \IEEEmembership{Senior Member, IEEE}

\thanks{Li Yang is with the Faculty of Business and Information Technology, Ontario Tech University, Oshawa, ON L1G 0C5, Canada (e-mail: li.yang@ontariotechu.ca)}%
\thanks{Mirna El Rajab and Abdallah Shami are with the Department of Electrical and Computer Engineering, Western University, London, ON N6A 3K7, Canada (e-mails: melrajab@uwo.ca, abdallah.shami@uwo.ca)}%
\thanks{Sami Muhaidat is with the Department of Electrical Engineering and Computer Science, Khalifa University, Abu Dhabi 127788, UAE (e-mail: sami.muhaidat@ku.ac.ae)}%
}

\markboth{Published in IEEE Transactions on Network and Service Management}
{}

\maketitle


\begin{abstract}
Zero-Touch Networks (ZTNs) represent a state-of-the-art paradigm shift towards fully automated and intelligent network management, enabling the automation and intelligence required to manage the complexity, scale, and dynamic nature of next-generation (6G) networks. ZTNs leverage Artificial Intelligence (AI) and Machine Learning (ML) to enhance operational efficiency, support intelligent decision-making, and ensure effective resource allocation. However, the implementation of ZTNs is subject to security challenges that need to be resolved to achieve their full potential. In particular, two critical challenges arise: the need for human expertise in developing AI/ML-based security mechanisms, and the threat of adversarial attacks targeting AI/ML models. In this survey paper, we provide a comprehensive review of current security issues in ZTNs, emphasizing the need for advanced AI/ML-based security mechanisms that require minimal human intervention and protect AI/ML models themselves. Furthermore, we explore the potential of Automated ML (AutoML) technologies in developing robust security solutions for ZTNs. Through case studies, we illustrate practical approaches to securing ZTNs against both conventional and AI/ML-specific threats, including the development of autonomous intrusion detection systems and strategies to combat Adversarial ML (AML) attacks. The paper concludes with a discussion of the future research directions for the development of ZTN security approaches.
\end{abstract}



\maketitle

\section{Introduction}

Future networks are envisioned to achieve fully autonomous network and service management, as human intervention is still required in current networks \cite{6g1}. Recently, several concepts or architectures have been proposed for network automation and optimization, such as Self-Organizing Network Management (SON), Intent-Based Network Management (IBN), and Autonomic Network Management (ANM) \cite{zsm2}. More recently, Zero-Touch Networks (ZTNs) have emerged as a transformative approach and a next-generation system for automating network operations to minimize manual network tasks, including network management, configuration, and optimization \cite{ztndf}. ZTNs aim to enhance network efficiency, reliability, and security by leveraging Artificial Intelligence (AI) and Machine Learning (ML) techniques, as well as the fifth generation of network (5G) technologies, like Software-Defined Networking (SDN) and Network Functions Virtualization (NFV). While ZTNs are centered on the automation of network operations, Zero-touch Network and Service Management (ZSM) proposed by the European Telecommunications Standards Institute (ETSI) \cite{zsm3} complements ZTNs by targeting the automation of network service lifecycle management. The networking and communication scientific community anticipates that AI/ML techniques will play a vital role in fully automating the management and orchestration of the sixth generation of networks (6G) \cite{6g1}.

AI/ML techniques have become key enablers for ZTNs and 5G/6G networks, realizing the rapid and accurate processing of massive volumes of network data to achieve network automation and optimization \cite{mythesis}. The implementation of ML approaches in network and service management leads to significant improvements in service efficiency, performance, and time management \cite{zsm1}. AI/ML methods can provide or enhance a wide range of network services and management functionalities, such as network behavior analysis, anomaly detection, traffic classification and forecasting, mobility prediction, and resource allocation.

In spite of the widespread adoption of AI/ML techniques in current networks, developing and deploying ML algorithms still requires extensive domain expertise and human effort \cite{automl1}. Additionally, traditional AI/ML models still have many limitations, such as human errors and insufficient adaptability. To address the limitations of traditional AI/ML models and realize ZTNs, Automated Machine Learning (AutoML) techniques emerge as potential solutions for network automation by enabling automated network data analytics for data-driven network services \cite{myautoml}. AutoML enables the automation of various data analytics and ML procedures, including automated data pre-processing, automated feature engineering, automated model selection, Hyper-Parameter Optimization (HPO), and automated model updating \cite{automl2}. Automated data preprocessing and feature engineering aim to enhance network data quality for improved analytics results, whereas automated model selection and HPO produce optimized ML models with optimal performance. Moreover, real-world networks typically operate in ever-changing environments, leading to model drift or data distribution changes. Automated model updating is a potential solution to address model drift issues and maintain the performance of AI/ML models \cite{myiotm}. Overall, AutoML techniques are effective solutions for realizing ZTNs.

Critical self-management functionalities in ZTNs include self-configuration, self-healing, self-optimization, and self-protection \cite{zsmsec1}. Among these functionalities, both self-healing and self-protection are particularly relevant to network security concerns, making network security mechanisms critical components in ZTNs. ZTN security exemplifies the concept that security should be an integral component of the network infrastructure \cite{zsmsec2}. By incorporating security measures into the network itself, management and maintenance are significantly simplified, since no additional hardware or software is required. 

Cybersecurity issues are primary challenges in ZTNs. Similar to general networks, ZTNs face various cyber threats, such as Denial of Service (DoS), eavesdropping, Man-in-the-Middle (MITM), web attacks, and malware. Apart from general network security threats, the adoption of ZTN technologies in Beyond 5G (B5G)/6G networks is envisioned to introduce additional security challenges and increased attack surfaces. These include threats to open Application Programming Interface (API), intent, closed-loop network automation, programmable network technology, and AI/ML models \cite{zsm1} \cite{zsmsec1}. Among these security threats, AI/ML-based attacks, or Adversarial Machine Learning (AML) attacks, are expected to introduce critical challenges in ZTNs/6G networks. AML attacks exploit the vulnerabilities of AI/ML models, causing them to make incorrect predictions/decisions and leading to security breaches and unauthorized access to sensitive data \cite{zsmsec3}. Therefore, it is crucial to develop effective security mechanisms that can detect, mitigate, and prevent security threats in ZTNs \cite{zsmsec2}.

\begin{table*}[htbp]
\caption{Summary of Existing Surveys on ZTN Security}
\centering
\scalebox{0.90}{
\begin{tabular}{|>{\centering\arraybackslash}m{8.0em}|>{\centering\arraybackslash}m{6.5em}|>{\centering\arraybackslash}m{6.5em}|>{\centering\arraybackslash}m{6.5em}|>{\centering\arraybackslash}m{8.5em}|>{\centering\arraybackslash}m{4.5em}|>{\centering\arraybackslash}m{5.2em}|>{\centering\arraybackslash}m{5.2em}|>{\centering\arraybackslash}m{4.5em}|}
\hline
\textbf{Paper} & \textbf{ZTN Security Vulnerabilities} & \textbf{AML Attacks for ZTNs} & \textbf{AI/ML Enablers for ZTNs} & \textbf{Automated Enabler for ZTNs (e.g., AutoML)} & \textbf{ML Case Study} & \textbf{AutoML Case Study} & \textbf{AML Defense Case Study} & \textbf{Open-Access Code} \\ \hline
Coronado \textit{et al.} \cite{zsm2} & \Checkmark & \Checkmark & \Checkmark & & & & & \\ \hline
Liyanage \textit{et al.} \cite{zsm1} & \Checkmark & \Checkmark & \Checkmark & & & & & \\ \hline
Benzaïd \textit{et al.} \cite{zsmsec1} & \Checkmark & \Checkmark & \Checkmark & & & & \Checkmark&  \\ \hline
Gallego-Madrid \textit{et al.} \cite{zsm4} & \Checkmark & & \Checkmark & & & & & \\ \hline
Ashraf \textit{et al.} \cite{zsm5} & \Checkmark & \Checkmark & \Checkmark & & & & & \\ \hline
Benzaïd \textit{et al.} \cite{zsm6} & \Checkmark & \Checkmark & \Checkmark & & \Checkmark & & & \\ \hline
Our Survey & \Checkmark & \Checkmark & \Checkmark & \Checkmark & \Checkmark & \Checkmark & \Checkmark & \Checkmark \\ \hline
\end{tabular}
\label{liter}
}
\end{table*}

\begin{figure*}[!t]
\centerline{
\includegraphics[width=11.5cm]{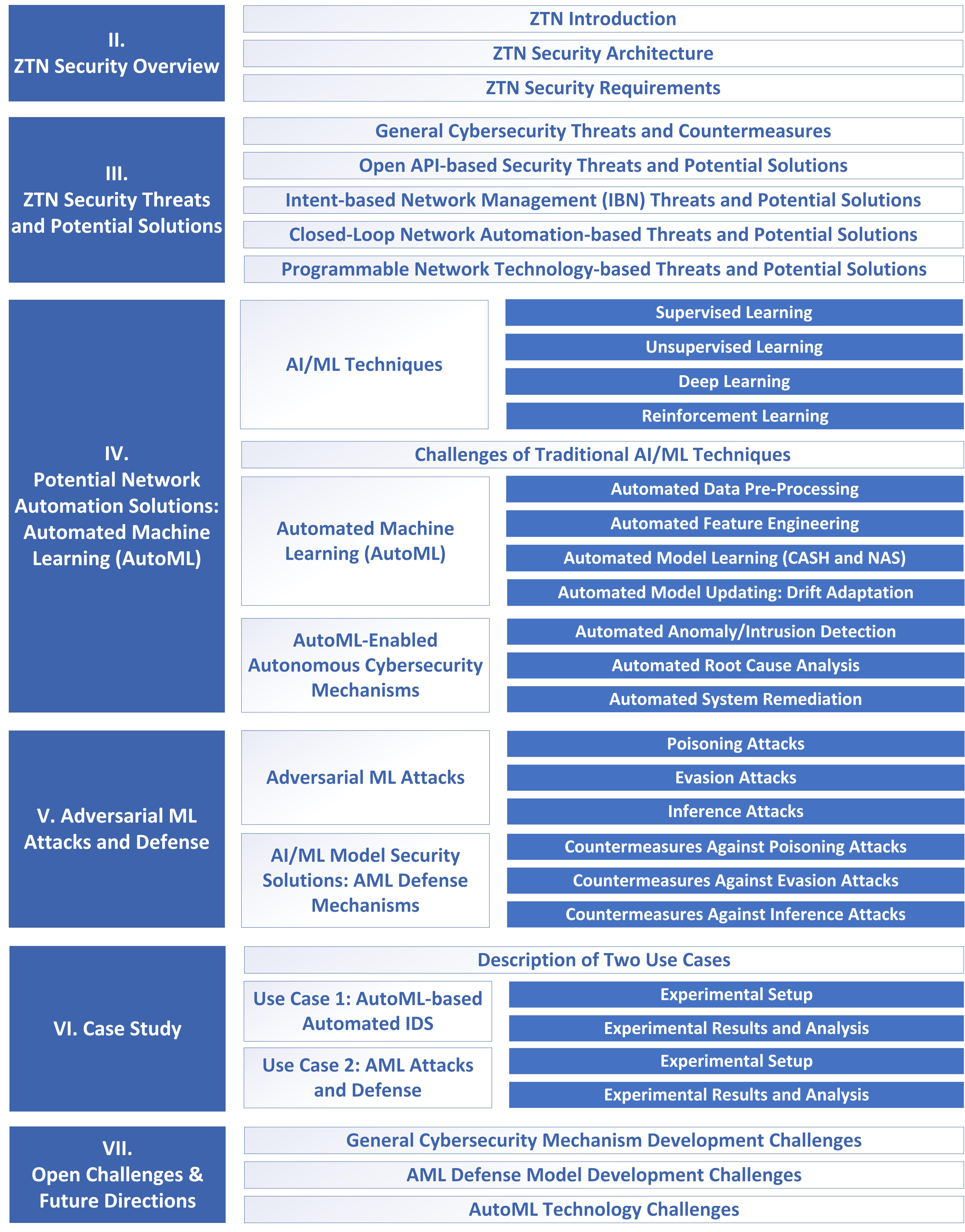}}
\caption{Overview of the survey's organization.}
\label{outline}
\end{figure*}

Although there is an increasing popularity in ZTN research, a comprehensive survey on ZTN security and cybersecurity automation has yet to be conducted. This survey paper aims to bridge the gap between ZTN \& network automation techniques and cybersecurity applications by reviewing existing challenges and potential solutions between these two domains. To the best of our knowledge, this survey constitutes the first comprehensive examination of ZTN security and automated cybersecurity methods, specifically through the exploration of AutoML techniques.

Specifically, this survey paper provides the following contributions:
\begin{enumerate}[label=\roman*)]
\item A comprehensive review of potential security issues and cyber-attacks in ZTNs, covering general cybersecurity challenges, security threats targeting ZTN services, and AI/ML model security issues. 
\item A detailed discussion of potential solutions to ZTN security issues, including general cybersecurity automation procedures, AutoML techniques, and AML defense mechanisms.
\item Two case studies that apply AutoML techniques to practical ZTN security tasks and conduct the cyber-defense exercise against AML attacks; the implementation code is publicly available on GitHub\footnote{
Code for this paper is available at: \url{https://github.com/Western-OC2-Lab/AutoML-and-Adversarial-Attack-Defense-for-Zero-Touch-Network-Security} }. 
\item A discussion of the open challenges and research directions of ZTN security methods. 
\end{enumerate}


The contributions of existing survey papers on ZTN security are summarized in Table \ref{liter}. Table \ref{liter} illustrates that all existing survey papers, including references \cite{zsm2}, \cite{zsm1}, \cite{zsmsec1}, \cite{zsm4}, \cite{zsm5}, and \cite{zsm6}, introduce ZTN security vulnerabilities and highlight AI/ML models as pivotal enablers for ZTNs. Among these, AML attacks are identified as critical threats to ZTN security in most surveys, with the exception of the paper \cite{zsm4}, which does not address AML attacks. Additionally, the paper \cite{zsmsec1} stands out as the sole survey providing an AML defense case study, and \cite{zsm6} is distinguished as the only work presenting a ML case study. Therefore, the primary unique contributions of this survey are contributions iii) and iv), where our paper is the first to comprehensively discuss automated security solutions for ZTNs, selecting AutoML as a potential solution, and providing comprehensive case studies on ML-based security, AutoML-based ZTN security, and AML attack and defense scenarios. Furthermore, this survey sets a precedent as the first ZTN security survey to share open-access code for the discussed case studies, making significant contributions to the field.

Therefore, this paper focuses on exploring AutoML techniques for automating cybersecurity mechanisms for ZTNs with network automation requirements and then discussing the safeguarding of AI/ML and AutoML models in ZTNs against AML attacks. Based on the two main focuses of this paper—automated cybersecurity solution development via AutoML techniques, and AML attack \& defense mechanisms—the structure of the remainder of this manuscript is outlined in Fig. \ref{outline} and as follows. Section \ref{S2} provides an overview of the ZTN security architecture and its requirements. Section \ref{S4} details the vulnerabilities and cyber-attacks targeting ZTNs. Section \ref{S5}, the first focus of this paper, introduces AI/ML \& AutoML techniques and discusses automating general security mechanisms with AutoML. Section \ref{S5-AML}, representing the second focus, delves into the most critical security threats to ZTNs and AI/ML models—AML attacks—and explores potential solutions. Section \ref{S6} features two case studies: the autonomous ZTN security framework utilizing AutoML techniques and the AML cyber-defense exercise. Section \ref{S7} discusses the challenges and future directions within the ZTN security and AutoML research domains. Finally, Section \ref{S8} concludes the paper.

\section{ZTN Security Overview} \label{S2}
\subsection{ZTN Introduction}

ZTNs have emerged as a solution for fully automating network operations, providing self-configuration, self-monitoring, self-healing, and self-optimization \cite{zsm4}. The adoption of ZTNs is driven by the demand for virtualized network functionality and the preference for software-based solutions over hardware-based alternatives \cite{zsm1}. They are designed to operate, conduct commercial activities, self-sustain, recover from disruptions, and perform other network-related tasks without human intervention. The deployability of ZTNs is facilitated by various APIs and End-to-End (E2E) programmability \cite{zsm1}. ZTNs are anticipated to play a crucial role in future networks, enabling the efficient distribution of network resources to meet the specific requirements of various industries and individuals \cite{zsm2}. The automation capabilities provided by ZTNs are essential for analyzing and responding to customer-specific needs, ensuring the desired Quality of Experience (QoE). In addition, ZTNs can evolve and upgrade through E2E automation, incorporating and updating new features that are required for practical applications.

ZTNs require a high level of automation, highlighting the importance of AI/ML techniques in transforming network management to achieve self-adaptation and self-reaction in highly dynamic environments. To facilitate network automation, Release 17 of the 3rd Generation Partnership Project (3GPP) includes a NFV/SDN approach via the Network Data Analytics Function (NWDAF) \cite{zsm5}. This integration allows for the collection and analysis of network data in real-time, enabling predictive maintenance, enhanced security, and optimized network performance. 
\subsection{ZTN Security Architecture} 
In the field of network and service management, cybersecurity automation can be achieved through the use of AI/ML techniques in combination with closed-loop operations \cite{zsmsec2}. Each management domain within the ZTN architecture contains a data analytics service that enables closed-loop network security and optimization operations. An example of a fully automated network security framework for ZTN architectures is proposed in the “Intelligent security and pervasive trust for 5G and beyond” (INSPIRE-5Gplus) project, funded by the European Union (EU) \cite{framework1}.

\begin{figure}[!t]
\centerline{
\includegraphics[width=7.8cm]{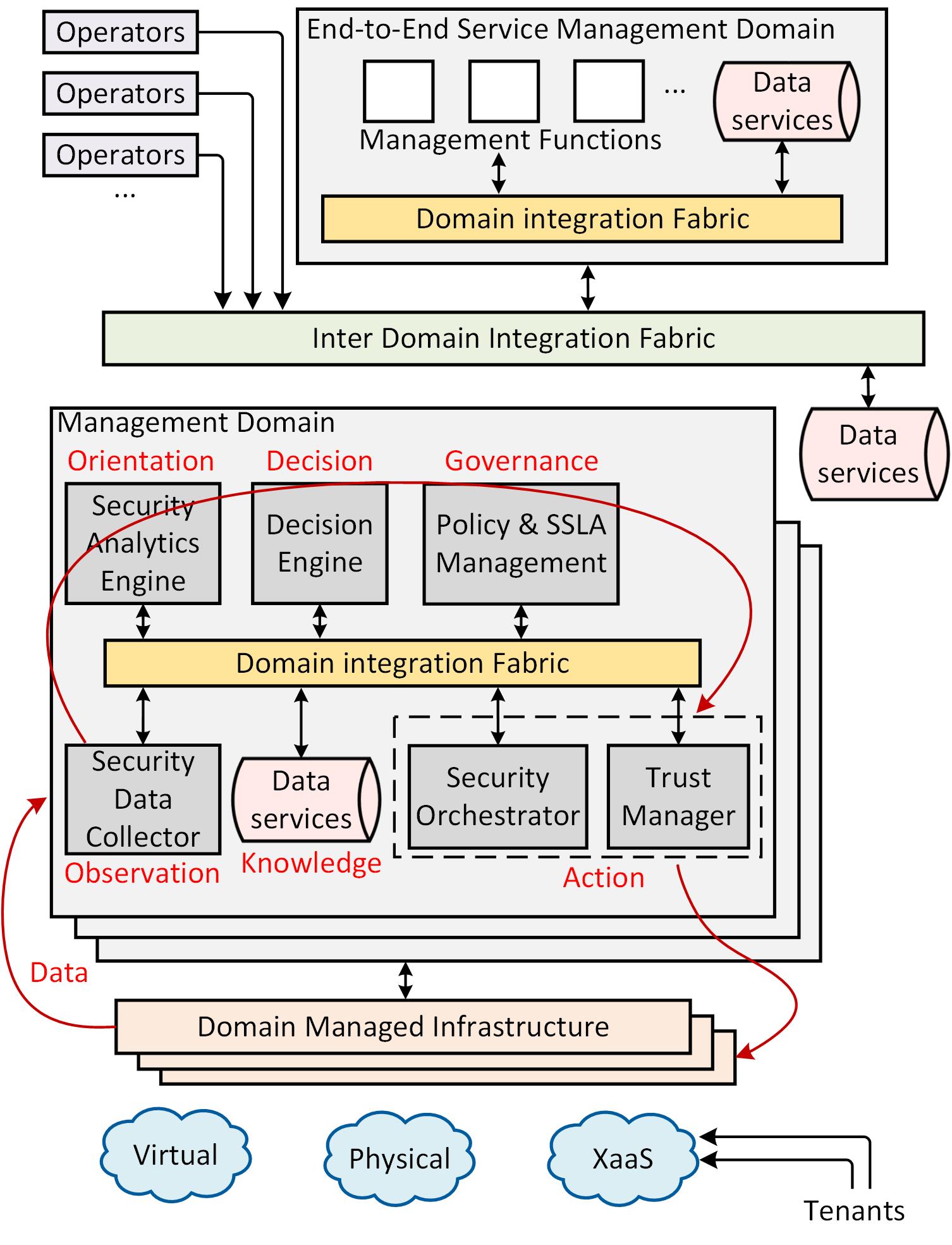}}
\caption{The ZTN security framework \cite{zsmsec2}.}
\label{zsm}
\end{figure}

Closed-loop operations are essential enablers of network automation within the ZTN architecture. Figure \ref{zsm} illustrates a domain-level security closed-loop ZSM framework, which consists of five stages: observation, orientation, decision, action, and learning \cite{zsmsec2} \cite{framework1}. In the first stage, data is observed and collected from various data sources by the security agents. The collected data is then oriented and analyzed by the security analytics service to comprehend its meaning and significance, which may involve data filtering and correlation analysis. In the subsequent stage, the decision engine uses AI/ML algorithms to recognize data patterns and make informed decisions or predictions for network optimization. Upon identifying the root causes and providing recommendations, the security orchestrator implements environment- and situation-appropriate actions or plans. Lastly, the system learns from the detected patterns and consequences to improve future reactions or decisions.  

E2E management domains are responsible for managing service and resource requests across different domains \cite{zsmsec2}. A management domain refers to a collection of services and resources organized based on various constraints, such as operations, functions, and deployment. Communication within domains is facilitated by the domain integration fabric, while communication between domains and E2E management domains is enabled by the inter-domain integration fabric. Within the ZTN architecture illustrated in Fig. \ref{zsm}, a management domain consists of a security data collector for data observation and collection, a security analytics engine for orientation, a decision engine enabled by AI/ML methods for decision-making, a policy and Security Service Level Agreement (SSLA) manager for governance, a security orchestrator and trust manager for action-taking, and data services for knowledge learning. 

\subsection{ZTN Security Requirements} 
The ZTN framework reference architecture plays an important role in the networking and cybersecurity landscape. As the world transitions toward more automated and intelligent networks, the need for stringent security measures becomes increasingly crucial. The ZTN security frameworks aim to reduce the risk of human error and create an efficient and secure network environment. For ZTN security frameworks, several essential security requirements must be met \cite{zsm1} \cite{zsmsec1} \cite{zsm6}: 
\begin{enumerate}[label=\roman*)]
\item \textit{Data Protection and Integrity}: The ZTN framework must support data protection mechanisms that cover data in use, in transit, and at rest. Ensuring the security of data management and data integrity is essential for ZTN frameworks. In addition, the ZTN framework must ensure the availability of network data, infrastructure resources, and management functions.
\item \textit{Privacy Preservation}: The ZTN framework should incorporate personal data privacy features that conform to the privacy-by-design and privacy-by-default principles. These principles ensure that privacy is built into the architecture from the outset, with the highest level of privacy settings applied by default and without user intervention.
\item \textit{Access Control and Security Policy Enforcement}: The ZTN framework should enable authenticated service users to authorize service access and follow security policies. This capability allows for the automated implementation of appropriate security policies based on the status of individual management services in relation to security requirements. Streamlining access control enhances overall security posture.
\item \textit{Intrusion Detection, Prevention, and Mitigation}: The ZTN framework must support automated detection, recognition, prevention, and mitigation of attacks and incidents. These capabilities are crucial for swiftly identifying and mitigating potential security threats, thereby minimizing the impact of cyber-attacks on the network and its services.
\item \textit{AI/ML Decision Supervision and Auditing}: To safeguard against vulnerabilities and cyber-attacks, the ZTN platform reference architecture should enable the supervision and auditing of privacy and security decisions made by AI/ML techniques. This mechanism guarantees that AI/ML-driven decisions establish security requirements, mitigate potential risks, and recover compromised devices while remaining resilient to AI/ML-based attacks.
\end{enumerate}

\begin{table*}[htbp]
\caption{Analysis of Security Threats in ZTNs and Their Countermeasures.}
\setlength\extrarowheight{1pt}
\begin{tabularx}{\textwidth}{|>{\centering\arraybackslash}p{2.5cm}|>{\centering\arraybackslash}p{2.8cm}|>{\centering\arraybackslash}X|>{\centering\arraybackslash}X|}
\hline
\textbf{Threat Category} & \textbf{Threat Examples} & \textbf{Description} & \textbf{Detailed Countermeasures} \\ \hline
General Network Threats & DoS/DDoS, Eavesdropping, MITM, SQL Injection, XSS & These foundational threats compromise network integrity, confidentiality, and availability, exploiting vulnerabilities across network layers and applications. & Deploy advanced IDS with AI/ML-driven anomaly detection, enforce end-to-end encryption, and utilize behavior analysis for early threat identification.\\ \hline
Open API-based Threats & Parameter Attacks, Identity Attacks, MITM Attacks, DDoS Attacks & As ZTNs increasingly rely on APIs for network operations, these threats exploit API vulnerabilities to manipulate or disrupt services. & Implement API security gateways, enforce strict authentication and authorization controls with OAuth2 and JWT, and use rate limiting along with comprehensive input validation to mitigate injection and overload attacks.\\ \hline
Intent-based Threats & Information Exposure, Undesirable Configuration, Abnormal Behavior & IBN vulnerabilities can lead to unauthorized access and system misconfigurations, undermining network security and efficiency. & Apply rigorous authentication mechanisms, secure intent transmission with TLS, and employ intent schema validation to prevent misconfigurations and ensure policy integrity.\\ \hline
Closed-Loop Network Automation-based Threats & MITM Attacks, Deception Attacks & Automated network configurations using AI/ML are susceptible to interception and manipulation, threatening the reliability of self-optimizing actions. & Secure AI/ML data pipelines, encrypt communication channels, and integrate real-time monitoring and anomaly detection to identify and mitigate threats swiftly.\\ \hline
Programmable Network Technology-based Threats & SDN Threats, NFV Threats & SDN and NFV introduce new security challenges, including unauthorized access and manipulation of network functions. & Strengthen SDN controllers and NFV orchestrators with multi-layered security measures, including role-based access control, traffic encryption, and the implementation of secure boot processes for virtual functions.\\ \hline
\end{tabularx}
\label{4-ztn}
\end{table*}

\section{ZTN Security Threats and Potential Solutions} \label{S4}
To ensure the cybersecurity of ZTNs, it is essential to implement effective E2E security management. The increasing number of connected devices and the expanding networks have resulted in a rapid increase in threats targeting 5G/6G networks. The threat surfaces of the ZTN architecture include open API threats, intent-based threats, closed-loop network automation-based threats, and programmable network technology-based attacks \cite{zsm1} \cite{zsm5}. This section discusses the primary security challenges specific to ZTNs, the ZTN architecture, and the corresponding potential counterattack mechanisms. The challenges presented in this section are summarized in Table \ref{4-ztn}. Moreover, the popularity of AI/ML models in enabling autonomous services in ZTNs has introduced a special challenge in the form of adversarial attacks, which aim to compromise AI/ML models and disrupt ZTN services \cite{zsm2} \cite{zsm5}. This challenge will be further addressed in Section \ref{S5-AML}.

\subsection{General Cybersecurity Threats and Countermeasures}
In the progression from 1G to 5G networks, technological advancements have enhanced communication capabilities but also escalated the spectrum of cybersecurity threats. Many existing vulnerabilities and cyber threats are still active in modern networks and can breach ZTNs. Key threats that have been identified across these network generations include Denial of Service (DoS) and Distributed DoS (DDoS) attacks \cite{mytnsm}, eavesdropping (e.g., port scan) \cite{mymth}, Man-in-the-Middle (MITM) attacks \cite{mitm}, web application attacks (notably Structured Query Language (SQL) injection and Cross-Site Scripting (XSS)) \cite{mymth}, and malware (e.g., viruses, worms, ransomware, spyware) \cite{mitm}. 

The persistence of these cyber threats necessitates the development of robust and advanced cybersecurity mechanisms to protect general networks and ZTNs. To defend against these cyber threats, comprehensive cybersecurity solutions incorporate four essential steps: anomaly detection, root cause analysis, system remediation, and intrusion prevention. The automation of these cybersecurity solutions is required in ZTNs, which are discussed in detail in Section \ref{S3-sec}.

\subsection{Open API-based Security Threats and Potential Solutions} 
Open APIs are crucial components for integrating web-based applications and facilitating communication within the ZTN framework, which plays an integral role in service provisioning, management, orchestration, and monitoring \cite{6gsec3} \cite{api1}. As 6G networks follow the trend set by 5G networks, open APIs are expected to play an even more significant role and be further enhanced \cite{6gsec1}. However, as APIs are vital to network operations, they also become attractive to cyber-attackers, posing various security threats to ZTNs \cite{zsm1}.

Cyber attacks exploiting open APIs can be divided into four major categories: parameter attacks, identity attacks, MITM attacks, and DDoS attacks \cite{zsm1} \cite{zsmsec1} \cite{6gsec1}:
\begin{enumerate}[label=\roman*)]
\item \textit{Parameter Attacks}: Parameter attacks take advantage of data transmitted through the API, such as query parameters, Hypertext Transfer Protocol (HTTP) headers, Uniform Resource Locators (URLs), and post content. Examples of parameter attacks include script insertions, SQL injections, buffer overflow attacks, and injection attacks targeting the common data services component of the ZTN framework \cite{zsm1}. These attacks can cause unauthorized data access, data manipulation, and DoS attacks.
\item \textit{Identity Attacks}: Identity attacks exploit vulnerabilities in the authentication and authorization procedures of ZTNs, such as using extracted API keys as credentials or delivering large quantities of random data to identify vulnerabilities \cite{6gsec1}. Insecure APIs can result in unauthorized access to domain orchestration services or E2E service orchestration services, allowing attackers to modify configurations or create E2E service instances to exhaust network resources \cite{zsm1}.
\item \textit{MITM Attacks}: MITM attacks occur when cyber-attackers intercept unencrypted API messages, capture confidential information, or manipulate messages \cite{zsmsec1}. These attacks can compromise the confidentiality and security of transmitted data.
\item \textit{DDoS Attacks}: APIs are also vulnerable to DoS and DDoS attacks, where attackers can overwhelm an API with a large number of requests, resulting in service unavailability \cite{zsm1}.
\end{enumerate}

API security is essential to restrict interactions with ZTN APIs to authorized users \cite{api2}. Several security measures, such as authorization, communication encryption, and input validation, can be implemented in ZTNs to mitigate API security threats \cite{zsmsec1}. The enhancement of authorization can be achieved through the application of Access Control Lists (ACLs), Role-Based Access Control (RBAC), and Attribute-Based Access Control (ABAC). These mechanisms are designed to aid in fulfilling the imperative of the least privilege \cite{zsmsec1}. To prevent MITM attacks, API messages must be encrypted, and Transport Layer Security (TLS) can be used to improve API message integrity and confidentiality. Validation of all input data in the URL, query parameters, HTTP headers, JavaScript Object Notation (JSON) schema, and content of ZTN APIs is essential for preventing injection attacks. Moreover, AI/ML techniques can be used to improve existing security mechanisms in ZTNs, such as automated API threat detection and mitigation.

\subsection{Intent-based Network Management (IBN) Threats and Potential Solutions} 
Intent-Based Network Management (IBN) offers a promising approach to integrate AI/ML techniques into 6G networks, addressing the limitations of traditional networks in terms of efficiency, flexibility, and security \cite{6gsec1}. By leveraging AI/ML techniques, IBN directly translates users' business intentions into network configurations, operations, and maintenance strategies. As part of the ZTN architecture, intent-based interfaces provide high-level abstractions and specify policies, enabling the decoupling of technology- and vendor-specific details \cite{api1}. However, IBN introduces security vulnerabilities, including information exposure, undesirable configuration, and abnormal behaviors \cite{6gsec1}:
\begin{enumerate}[label=\roman*)]
\item \textit{Information Exposure}: Intent-based interfaces may expose confidential information about the application's desires, such as connecting with peers, advertising services or content, and controlling network traffic \cite{zsmsec1}. This information can be intercepted without authorization, compromising system security objectives (\textit{e.g.}, privacy and confidentiality) and opening the backdoor to additional attacks \cite{zsm1}.
\item \textit{Undesirable Configuration}: Malicious modification of an intent-based service model may result in a configuration of an undesirable security level, leaving the network segment vulnerable to security threats \cite{zsmsec1}. For instance, a cyber-attacker could alter a high-security-level intent to a low-security-level intent, resulting in deficient security measures in a network segment.
\item \textit{Abnormal Behavior}: A malformed intent sent to a domain orchestration service may trigger abnormal behavior, including service disruption or a DoS attack \cite{zsmsec1}. Alternatively, the service may transfer the intent to a policy, resulting in a misconfigured network that causes network outages or security breaches.
\end{enumerate}

To mitigate these intent-based threats, several potential mechanisms can be employed \cite{zsmsec1}:
\begin{enumerate}[label=\roman*)]
\item \textit{Authentication and Authorization}: OpenID Connect, signed JSON Web Tokens (JWT), and OAuth2.0 can ensure mutual authentication between intent producers and consumers and restricted access to intent-based interfaces.
\item \textit{Secure Transport Protocol}: The exchange of intents should occur over a secure transport protocol, such as TLS 1.2, to maintain intent integrity and confidentiality, thereby preventing intent tampering and eavesdropping.
\item \textit{Intent Validation and Conflict Resolution}: The intent engine should be able to validate the intent format and detect/resolve potential conflict situations, mitigating the impact of malformed or conflicting intents on insufficient network operation and security.
\end{enumerate}

Intent-based interfaces can contribute to the overall security and efficacy of ZTNs within the context of 6G mobile networks by addressing these security concerns and implementing the suggested mitigation mechanisms.

\subsection{Closed-Loop Network Automation-based Threats and Potential Solutions} 
6G networks will leverage closed-loop network automation for zero-touch management capabilities \cite{cla1}. Closed-Loop Automation (CLA) continuously monitors, identifies, adjusts, and optimizes network performance to enable self-optimization \cite{6gsec1} \cite{zsm1}. As 6G technologies evolve, external CLA capabilities become necessary for addressing the expanding threat landscape. Using AI/ML techniques, CLA security mechanisms have the potential to autonomously detect and rapidly mitigate threats. 

However, several security threats are driven by closed-loop networked automation, such as MITM, deception attacks, and other unknown threats \cite{zsm1} \cite{zsmsec1}. In MITM attacks, an intruder intercepts communications between two parties to eavesdrop or manipulate traffic remotely. The cyber-attacker obtains user credentials and confidential information, and they may interrupt message exchanges, resulting in data corruption. Deception attacks rely on manipulating the target to act based on false information, convincing them to perceive an incorrect version of the truth as fact, and causing the propagation of false feedback information. 

To address security threats in CLA, implementing secure communication channels between network components can reduce the risk of MITM and deception attacks and prevent unauthorized access \cite{zsmsec1}. Additionally, AI/ML techniques can be used to monitor network traffic and detect anomalies for real-time CLA attack detection. Validation and integrity tests at various phases of a closed-loop process can also assist in preventing the injection and manipulation of malicious data. 

\subsection{Programmable Network Technology-based Threats and Potential Solutions} 
ZTNs leverage SDN and NFV technologies to construct programmable networking solutions \cite{zsm1} \cite{zsm5}. Despite their potential benefits, these technologies introduce new attack vectors and security risks that compromise the application, data, and control layers of SDN \cite{sdn1}.
\subsubsection{SDN Threats and Potential Solutions}
The Open Network Foundation (ONF) has identified various threats to SDN, including spoofing, tampering, repudiation, information leakage, DoS, and privilege escalation \cite{zsmsec1}. These threats can target various SDN layers (\textit{i.e.}, the control, data, and application layers) and often result from design and implementation limitations in existing SDN controller platforms \cite{sdn2}. The risks are increased when users have programmatic access to SDN, particularly when they use third-party applications or standard-based solutions to access the network. Insufficient isolation can lead to cyber-attacks such as network manipulation and Address Resolution Protocol (ARP) spoofing \cite{zsm1} \cite{zsm5}.
Several solutions have been suggested for mitigating SDN-related attacks \cite{zsmsec1} \cite{sdn2}. These include:
\begin{enumerate}[label=\roman*)]
\item Implementing authorization mechanisms such as Role-Based Access Control (RBAC) to regulate access levels and prevent privilege abuse.
\item Ensuring the confidentiality and integrity of messages exchanged on communication channels, both in the North-South and East-West interfaces, through the use of encryption, digital signature, and Message Authentication Code (MAC) algorithms.
\item Utilizing tamper-proof devices like Trusted Platform Modules (TPM) to secure sensitive data, encryption keys, passwords, and certificates. 
\item Implementing a distributed SDN controller architecture to defend against DoS/DDoS attacks and ensure system availability.
\end{enumerate}

\subsubsection{NFV Threats and Potential Solutions}
NFV is susceptible to various virtualization and networking threats, as well as threats resulting from their combination \cite{nfv1}. These threats include generic virtualization threats like memory leakage and interrupt isolation, generic networking threats like flooding attacks and routing security issues, and threats that arise from the integration of virtualization and networking, such as introspection attacks. Virtual Network Functions (VNFs) are vulnerable to design, implementation, and configuration weaknesses resulting in inappropriate monitoring of data that misleads ZTN service intelligence and E2E analytics \cite{zsm1}.
The NFV Security Group of ETSI has proposed various potential security mechanisms to counter NFV-related threats \cite{zsmsec1} \cite{nfv1}:
\begin{enumerate}[label=\roman*)]
\item \textit{Trusted Platform Module (TPM) and Virtual TPM}: These are recommended for ensuring the integrity of VNF booting and enabling remote verification.
\item \textit{Trusted Execution Environments}: Examples include Intel's Software Guard Extensions (SGX) and Secure Encrypted Virtualization (SEV) for Advanced Micro Devices (AMD). These are utilized for establishing the confidentiality and integrity of VNFs at runtime.
\item \textit{Traffic Monitoring and Filtering}: Tools such as Intrusion Detection Systems (IDSs) and firewalls are used for detecting DoS/DDoS attacks.
\item \textit{Robust Authentication and Authorization Mechanisms}: These are crucial for preventing NFV Management and Orchestration (MANO) hijacking.
\item \textit{Secure Communication Channels}: These are recommended for preventing MITM attacks.
\item \textit{System Hardening Techniques}: These can help prevent the exploitation of software vulnerabilities in NFV components.
\end{enumerate}

In essence, the security framework of ZTNs faces a broad spectrum of threats targeting open APIs, intent-based networking configurations, automated network operations, and advanced network programming functionalities. These threats compromise not just the core principles of security—integrity, confidentiality, and availability—but also put to test the flexibility and operational efficacy of ZTN architectures. Addressing these challenges requires the deployment of a multifaceted security strategy. This includes enforcing stringent access control, implementing encryption standards, employing advanced detection systems, and ensuring rigorous validation of inputs. Through such proactive measures, ZTNs can bolster their defenses against complex cyber threats, thereby maintaining a secure and dependable operational environment amidst the dynamic landscape of cybersecurity threats. This approach highlights the critical need for ongoing evaluation and enhancement of security protocols to safeguard ZTNs against existing and emerging cyber risks. Moreover, given the high-level network automation requirements of ZTNs, autonomous cybersecurity methods are expected to enhance ZTN security. AutoML techniques emerge as potential solutions for autonomous cybersecurity solutions in ZTNs, and are introduced in the next section.

\section{Potential Network Automation Solutions: Automated Machine Learning (AutoML)} \label{S5}

\subsection{AI/ML Techniques} \label{S5-ML}
AI/ML techniques are becoming increasingly vital for strengthening 6G security. Traditional security solutions, such as firewalls and conventional IDSs, struggle against complex cyber-attacks within a 6G environment. AI/ML techniques have proven to be a viable solution for enhancing security across multiple network layers, spanning from the physical layer to the network layer \cite{6gsec2}. 

At the physical layer, AI/ML significantly bolsters security defenses by improving the performance of detection engines, mitigating vulnerabilities such as eavesdropping and jamming attacks. On the network layer, AI/ML has been leveraged in several key technologies to enhance system performance within the 6G landscape. These include verifying node behavior to detect insider attacks, predicting potential network attacks, optimizing radio and computing control policies, determining the prioritization of equipment recovery, and scrutinizing traffic or network access behavior \cite{6gsec2}.

AI/ML plays a crucial role in enabling efficient ZTN management within 6G networks by automating and optimizing a wide range of network operations. ML techniques can be categorized into four groups, each with its specific applications in the context of ZTN services functionalities \cite{zsm4} \cite{ml1}.

\subsubsection{Supervised Learning}
Supervised Learning (SL) techniques are frequently utilized in ZTNs for the automation of network and service management \cite{ml2}. These techniques are especially effective in classification and regression problems, where the objective is to predict a specific class or output value given a particular input. SL is employed for tasks such as traffic classification, service requirement prediction, and user behavior forecasting. There are many different types of SL algorithms, such as linear regression, logistic regression, decision trees, random forests, XGBoost, LightGBM, K-Nearest Neighbors (KNN), Support Vector Machines (SVM), naive Bayes, and Neural Networks (NN) \cite{myautoml} \cite{mylccde} \cite{myguelph}.
\subsubsection{Unsupervised Learning}
Unsupervised Learning (UL) techniques are essential in understanding and organizing data without the need for labeled examples. They are used in ZTNs to group network traffic flows with similar characteristics and assign them to appropriate network slices. This ensures optimal utilization of network resources and allows for differentiated levels of Quality of Service (QoS). Commonly employed UL algorithms for these purposes encompass K-means, Gaussian mixture models, and Principal Component Analysis (PCA) \cite{myhpo} \cite{ml3}.
\subsubsection{Deep Learning}
Deep Learning (DL) techniques, based on artificial neural networks, address the challenges posed by the necessity for manual feature extraction in traditional ML approaches \cite{mycnn} \cite{dl1}. They enable the efficient processing of raw natural data and facilitate the implementation of large state-action spaces. Common DL algorithms include Multi-Layer Perceptron (MLP), Convolutional Neural Networks (CNNs), and Recurrent Neural Networks (RNNs) \cite{myhpo}. In the context of ZTNs, DL algorithms such as deep Q-learning and Soft Actor-Critic (SAC) are used for tasks like traffic classification and prediction, Radio Access Network (RAN) management, multi-domain orchestration, and resource management.
\subsubsection{Reinforcement Learning}
RL techniques enable adaptive decision-making in network management by modeling problems as driven by rewards and punishments \cite{rl2}. They find use in various applications, including slice admission strategies, data migration in Multi-access Edge Computing (MEC), Radio Access Technology (RAT) selection, and resource allocation for network services \cite{zsm4}. RL algorithms are categorized as model-based and model-free models, and examples such as proactive resource allocation and Q-learning are utilized in ZTN scenarios.
\subsection{Challenges of Traditional AI/ML Techniques}
Despite the widespread use of traditional ML algorithms for various cybersecurity tasks, they encounter several critical challenges \cite{automl1} \cite{myautoml} \cite{automl2}:
\begin{enumerate}[label=\roman*)]
\item \textit{Time-Consuming Model Development}: Traditional ML and data analytics techniques involve a sequence of manual tasks, such as data pre-processing, feature engineering, model selection, and hyperparameter tuning. These tasks require significant time and effort from ML developers and researchers, which can impede the rapid development and deployment of ML models within 5G/6G networks. 
\item \textit{Human Bias and Errors}: The inherent subjectivity involved in manual ML model development can introduce potential biases and errors. These can result in issues such as overfitting and underfitting, leading to a degradation in model performance. 
\item \textit{Expertise Requirements}: The deployment of traditional ML algorithms in practical applications often necessitates a high level of expertise. This can create a barrier to interdisciplinary collaboration by increasing the demand for skilled ML professionals.
\item \textit{Insufficient Adaptability}: Traditional ML models are often slow or unable to react to ever-changing data streams in dynamic networking environments. They generally lack the ability to detect unknown or zero-day attacks.
\end{enumerate}

While traditional ML algorithms present certain challenges, they remain vital techniques in the pursuit of enhanced security within the landscape of 6G networks. By addressing these challenges, we can further leverage the potential of AI/ML in cybersecurity, improving the resilience and adaptability of our future networks.

\subsection{Automated Machine Learning (AutoML)} \label{S5-AutoML}
Although ML algorithms can learn data without human intervention, there are certain procedures in ML pipelines that require experienced data scientists. These procedures include preparing appropriate and sanitized data, selecting the most suitable ML algorithm, tuning hyperparameters, and determining whether the ML model needs updating. Data scientists usually experiment with various pre-processing methods and ML algorithms to find the optimal combination. These procedures are human-centric, time-consuming, and require specialized expertise in ML and data analysis \cite{myautoml}. Automating the process of designing and tuning ML methods is known as AutoML. AutoML refers to the fully automated process of applying ML algorithms to real-world practical applications, allowing both beginners and experts to develop ML models in a more efficient way. 

AutoML aims to enable domain specialists to create ML models automatically without requiring extensive knowledge and experience in ML. AutoML addresses the shortage of data scientists and has the potential to significantly improve the performance and effectiveness of ML models by shortening work cycles and increasing model performance \cite{myautoml}. Given the high demand for network automation, AutoML techniques have become necessary components in ZTNs networks. 

ZTNs represent autonomous networks with self-healing and self-protection capabilities based on the data collected and analyzed across all network activities \cite{zsmsec1}. Robust and powerful AI/ML models are key components in safeguarding ZTNs by transforming conventional network security operations into automated and intelligent operations. In the general cybersecurity process discussed in Section \ref{S3-sec}, supervised and unsupervised AutoML techniques can be developed to automatically detect malicious cyber-attacks. RL-based AutoML techniques can then be used to automatically determine appropriate actions to defend against cyber-attacks and aid in system remediation.

\begin{figure*}[!t]
\centerline{
\includegraphics[width=14.5cm]{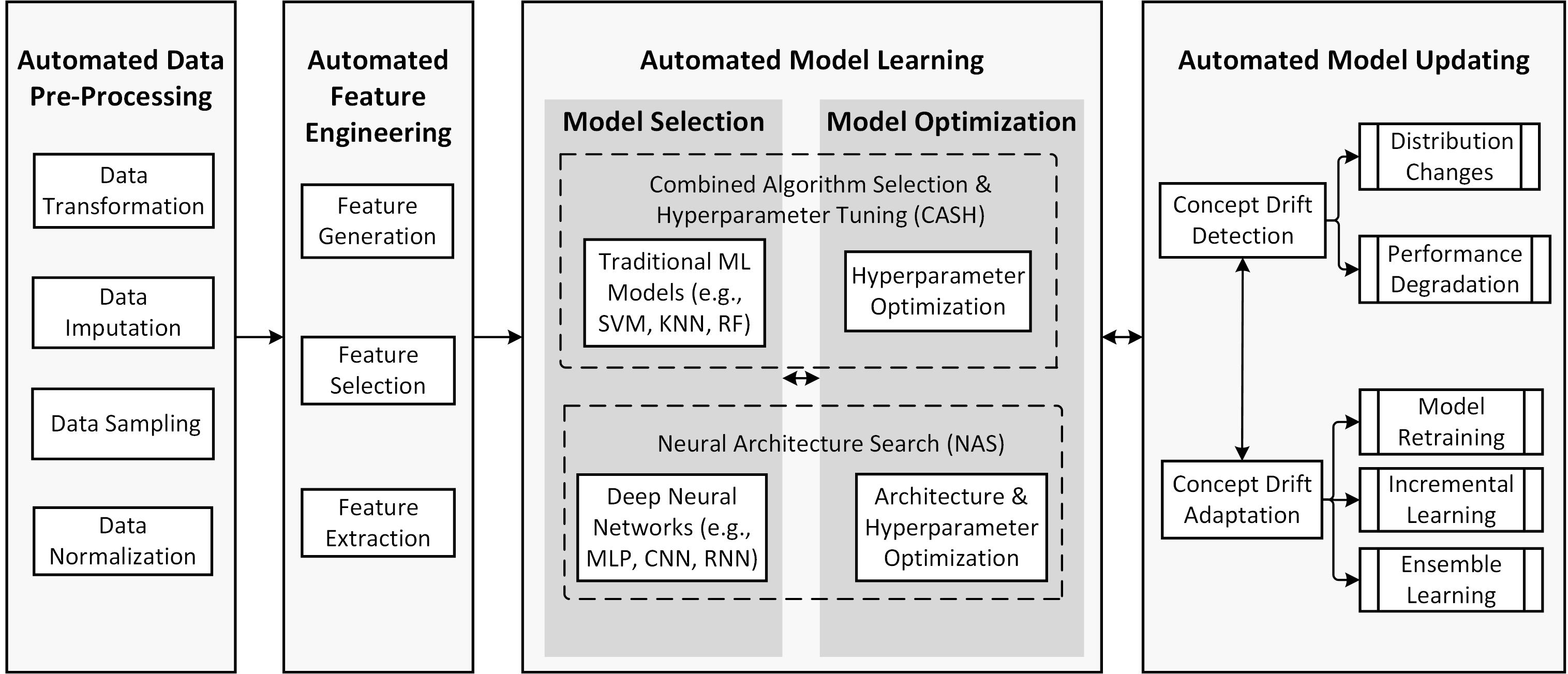}}
\caption{An overview of the general AutoML framework.}
\label{automl}
\end{figure*}

Figure \ref{automl} illustrates the general AutoML framework, which consists of four stages: automated data pre-processing, automated feature engineering, automated model learning, and automated model updating. The specifics of each stage of the AutoML pipeline are described below.

\subsubsection{Automated Data Pre-Processing}
Data pre-processing enhances data quality for ML model development by addressing common issues such as outliers, missing values, and class imbalance \cite{automl3}. Automated Data Pre-processing (AutoDP) is an essential component of AutoML, streamlining the time-consuming and tedious pre-processing stage. AutoDP tasks can be categorized into four main areas: data transformation, imputation, balancing, and normalization \cite{myautoml} \cite{automl4}.
\begin{enumerate}[label=\roman*)]
\item \textit{Data Transformation}: It involves encoding techniques for converting categorical features into continuous ones (\textit{e.g.}, label encoding and one-hot encoding) and discretization techniques for transforming continuous features into categorical ones \cite{myautoml}.
\item \textit{Data Imputation}: It addresses the issue of missing values commonly found in real-world network datasets, which can significantly impact the learning performance of ML models. Common data imputation approaches include zero/mean/median imputation for numerical and categorical features, forward/backward filling for time-series data, and ML model-based imputation techniques for advanced imputation \cite{impute}.
\item \textit{Data Balancing}: It addresses class imbalance issues in network datasets for classification problems, which can result in ML model performance degradation. Resampling techniques, such as over-sampling techniques (\textit{e.g.}, the Synthetic Minority Over-sampling TEchnique (SMOTE) \cite{mymth} and ADAptive SYNthetic (ADASYN) \cite{adasyn}) and under-sampling methods (\textit{e.g.}, random under-sampling), can be employed to mitigate class imbalance by generating minority class samples or removing majority class samples \cite{balance}.
\item \textit{Data Normalization}: It ensures fair treatment of features with different scales by ML models, especially when distance calculations are involved in model learning. Z-score and min-max normalization are two common approaches used in ML model learning, with the choice depending on the occurrence and percentage of outliers \cite{norm}.
\end{enumerate}

\subsubsection{Automated Feature Engineering}
Feature Engineering (FE) is a critical aspect of the ML pipeline that transforms or creates features from existing data to improve data quality. FE can determine the upper limit of ML applications \cite{automl2} \cite{automl3}. However, manual FE is laborious and requires domain knowledge. To address this issue, Automated FE (AutoFE), an important AutoML procedure, automates the generation and selection of relevant features, making the process more efficient and reproducible \cite{myautoml}.
AutoFE combines three main components: feature generation, feature selection, and feature extraction \cite{myautoml}. The generate-and-select strategy is employed by state-of-the-art AutoFE approaches to automate the traditional FE process, such as the AutoFE process used in the AutoFeat \cite{autofeat} library. AutoFeat \cite{autofeat} is a Python library that provides a multi-step AutoFE process based on the Lasso regression model to improve the traditional FE process. The general AutoFE process involves creating an exhaustive feature pool through feature generation methods and selecting or extracting valuable features using feature selection and extraction methods \cite{autofe}.
\begin{enumerate}[label=\roman*)]
\item \textit{Feature Generation}: It aims to create new features by transforming or combining existing features to improve the generalizability and robustness of ML models. Common feature generation operations include unary operations (\textit{e.g.}, logarithmic operations), binary operations (\textit{e.g.}, addition, subtraction, multiplication,  and division), and high-order operations (\textit{e.g.}, maximum, minimum, and average) \cite{automl5}. For network time-series data, features such as date/time features, lag features, and window-based features can be generated to improve model performance.
\item \textit{Feature Selection}: It aims to identify the most appropriate features and remove irrelevant or redundant ones. Automated feature selection, or AutoFS, frames the feature selection problem as an optimization problem to automatically determine the optimal number or percentage of selected features \cite{autofs}. Existing feature selection methods can be categorized into filter methods (\textit{e.g.}, Information Gain (IG), and Pearson correlation), wrapper methods (\textit{e.g.}, Recursive Feature Elimination (RFE)), and embedded methods (\textit{e.g.}, Lasso regularization and decision tree-based algorithms) \cite{myautoml} \cite{mymth} \cite{rfe}.
\item \textit{Feature Extraction}: It aims to reduce dimensionality by applying mapping functions, creating a more concise representation of the original dataset while potentially increasing model learning efficiency \cite{fealex}. Common feature extraction methods include Principal Component Analysis (PCA), Linear Discriminant Analysis (LDA), and Autoencoders (AE) \cite{automl2}. Feature extraction is typically utilized when the feature set remains highly dimensional or under-performing after feature generation and selection.
\end{enumerate}

In summary, AutoFE is a dynamic combination of feature generation, feature selection, and feature extraction to automate the FE process for improved efficiency and reproducibility. More accurate and efficient ML models can be developed by selecting the appropriate AutoFE methods and optimization techniques, ultimately advancing network performance.

\subsubsection{Automated Model Learning (CASH and NAS)}
The primary challenge in applying ML algorithms to real-world tasks resides in the selection and configuration of the most appropriate ML model to achieve optimal results. For a prediction task, the accuracy of different models with different configurations may vary significantly \cite{automl3}. As a result, it is crucial to identify the most suitable ML model with optimal hyperparameter configuration.

Hyperparameters are the parameters of ML algorithms that define the architecture of ML models and must be specified before model learning \cite{myhpo}. According to the domains of hyperparameters, they can be classified as continuous hyperparameters (\textit{e.g.}, the learning rate of neural networks), discrete hyperparameters (\textit{e.g.}, number of nearest neighbors in KNN), and categorical hyperparameters (\textit{e.g.}, the kernel type in SVM). Additionally, certain hyper-parameter configurations include conditionality. For example, in neural networks, the number of hidden layers and the number of neurons in each layer have strong correlations \cite{myhpo}. Conditional hyper-parameters must be tuned together to identify the optimal configuration.

Selecting an appropriate ML algorithm and hyperparameter values can be seen as a search problem. The set of potential ML models and their corresponding hyperparameter combinations constitute a search space, wherein each ML model, along with its specific hyperparameter configuration, represents a point within this space. Detecting the optimal point in the search space is a global optimization problem \cite{myhpo}.

Therefore, using optimization techniques to automatically detect the optimal ML algorithm and hyperparameter configuration is defined as Combined Algorithm Selection and Hyper-parameter (CASH) optimization problems \cite{cash}. CASH systems consist of two levels: automated model selection and Hyper-Parameter Optimization (HPO). At the first level, the suitable ML models are selected based on their default hyperparameters. At the second level, model-specific hyperparameters are optimized to obtain the final model with the best performance \cite{automl3}. 

Generally, CASH is defined as identifying the most suitable ML model $\mathcal{A}^{\star}$ with its optimal hyperparameter configuration ${\lambda}^{\star}$ that minimizes the loss function $\mathcal{L}$, as depicted by the following equation: 
 \begin{equation}
\mathcal{A}^{\star}, {\lambda}^{\star} \in \underset{\mathcal{A}^{(j)} \in \mathcal{A}, \lambda^{(j)} \in \boldsymbol{\lambda}}{\operatorname{argmin}} \frac{1}{K} \sum_{i=1}^{K} \mathcal{L}\left(\mathcal{A}^{(j)},  \lambda^{(j)}, D_{\text {train}}^{(i)}, D_{\text {valid}}^{(i)}\right)
\end{equation}
where $\mathcal{A}^{(j)} \in \mathcal{A}$ is the ML algorithm to be selected, $\lambda^{(j)} \in \boldsymbol{\lambda}$ refers to the hyperparameters to be optimized, $D_{train}$ and $D_{valid}$ are the training and validation datasets, and $K$ indicates the number of folds in k-fold cross-validation.

Similarly, Neural Architecture Search (NAS) is the process of automating the design of DL models \cite{automl1}. Apart from the hyperparameters in DL models, such as the learning rate, NAS techniques focus on automatically determining the optimal neural network architecture, such as the number of hidden layers, the type of each layer, and the number of neurons in each layer. 

Optimization techniques are used to solve CASH and NAS problems by automatically detecting the most suitable ML model or DL architecture with the best-performing hyperparameter configuration as the optimal solution \cite{myhpo}. Common optimization techniques for CASH and NAS tasks include Grid Search (GS), Random Search (RS), Bayesian Optimization with Gaussian Process (BO-GP), Bayesian Optimization with Tree Parzen Estimator (BO-TPE), Hyperband, Genetic Algorithm (GA), Particle Swarm Optimization (PSO), and Reinforcement Learning (RL) \cite{myautoml}. Table \ref{5-cash} provides a summary of these optimization techniques, outlining their descriptions, strengths, and limitations. Suitable optimization methods can be selected for specific tasks according to their respective strengths and limitations. 

\begin{table*}[htbp]
\caption{The comparison of optimization methods for CASH and NAS tasks.}
\setlength\extrarowheight{1pt}
\centering
\begin{tabular}{|>{\centering\arraybackslash}p{2.8cm}|>{\raggedright\arraybackslash}p{5.5cm}|>{\raggedright\arraybackslash}p{4cm}|>{\raggedright\arraybackslash}p{4cm}|}
\hline
\textbf{Optimization Method} & \textbf{Description} & \textbf{Strengths} & \textbf{Limitations} \\ 
\hline
{Grid Search (GS)} & A brute-force search method that evaluates all the given hyperparameter values. & · Straightforward to implement. \newline · Works well with categorical hyperparameters only. & · High computational cost. \\ 
\hline
{Random Search (RS)} & Randomly samples hyperparameter configurations from the search space until the specified budget is exhausted. & · Faster than GS. \newline · Allows for parallel execution. & · Does not take previous outcomes into account. \newline · Inefficient for conditional hyperparameters. \\ 
\hline
{Bayesian Optimization with Gaussian Process (BO-GP)} & Selects future hyperparameter configurations based on previous evaluations modeled by the Gaussian process. & · Rapid convergence for continuous hyperparameters. & · Inefficient for other types of hyperparameters. \\ 
\hline
{Bayesian Optimization with Tree Parzen Estimator (BO-TPE)} & Selects future hyperparameter configurations based on previous evaluations modeled by the tree-Parzen estimator. & · Effective for all types of hyperparameters. & · Limited parallelization capabilities. \\ 
\hline
{Hyperband} & Trains on randomly generated subsets and reduces the computational cost by balancing evaluation cost and accuracy. & · Supports parallel execution. & · Inefficient for conditional hyperparameters. \newline · Requires small budget subsets to be representative. \\ 
\hline
{Genetic Algorithm (GA)} & A population-based heuristic search technique that passes the well-performing hyperparameter values to future generations. & · Works well with all types of hyperparameters. \newline · Does not require optimal initialization. & · Limited parallelization capabilities. \\ 
\hline
{Particle Swarm Optimization (PSO)} & Enables communication and cooperation among particles to gradually identify the global optimum. & · Effective for all types of hyperparameters. \newline · Enables parallel execution. & · Requires appropriate initialization. \\
\hline
{Reinforcement Learning (RL)} & Efficiently explores the search space and adapts its strategy based on the performance feedback. & · Performs well with NAS tasks for DL models. & · High computational cost. \\
\hline
\end{tabular}
\label{5-cash}%
\end{table*}

\subsubsection{Automated Model Updating: Drift Adaptation} \label{AMU}
After the automated model learning process to train optimal ML models, these models are deployed into real-world production environments, leveraging their capabilities to solve complex tasks. However, when these ML models are introduced into dynamic networking systems, they may encounter substantial performance degradation due to model drift or concept drift issues \cite{myiotm}. 

Model drift refers to the decline in the ML model’s performance over time, causing model reliability issues in dynamic environments. Model drift can be further classified as data drift and concept drift. Data drift refers to the unpredictable changes in data distributions, while concept drift refers to the shifts in the relationships between the input features and the target variables \cite{drift1}. In cybersecurity applications, changes in the benign event data distributions are examples of data drift, while zero-day attacks exemplify concept drift \cite{myautoml}. Both data and concept drift can lead to the degradation of ML model performance, thereby underscoring the necessity for automated model updating. This process aims to preserve high prediction accuracy in non-stationary networking environments \cite{myautoml}. This process aligns with continual learning, a crucial ML procedure for data stream analytics.

To handle model drift and ensure ML models' continuous reliability, learning systems must promptly detect and adapt to these changes. Drift detection and adaptation methods are essential components of automated model updating designed for streaming data with data or concept drift issues. Drift detection methods are primarily classified into two categories \cite{mytii}:
\begin{enumerate}[label=\roman*)]
\item \textit{Distribution-based Methods}: They detect drift by monitoring data drift or data distribution changes in time windows. Techniques such as ADaptive WINdowing (ADWIN) \cite{adwin}, Information Entropy (IE), and Kullback-Leibler (KL) divergence \cite{kl} are commonly used for distribution-based drift detection. These methods are often used in systems with limited memory and can provide high interpretability. However, they may incur higher computational costs than performance-based methods.
\item \textit{Performance-based Methods}: They track changes in performance or prediction error rates of ML models to identify model drift. Popular methods like Drift Detection Method (DDM) and Early Drift Detection Method (EDDM) use pre-defined thresholds to determine the occurrence of drift \cite{mypwpae}. Performance-based methods are particularly effective for detecting sudden drift in data streams but may have limitations in identifying gradual drift.
\end{enumerate}

Once a drift is detected, adapting to the drift becomes crucial to updating the ML models and maintaining their performance in the context of data streams. There are three primary types of drift adaptation methods \cite{myautoml}: 
\begin{enumerate}[label=\roman*)]
\item \textit{Model Retraining}: Model retraining is a simple approach for drift adaptation that involves updating learning models on newly arrived data streams. Strategies include full retraining, partial retraining, and instance weighting. The full retraining process entails retraining the learning model on all available samples in the dataset. Partial retraining methods, such as Optimized Adaptive and Sliding Windowing (OASW) \cite{myiotm} and KNN-ADWIN \cite{knnadwin}, use adaptive or sliding windows to detect model drift and collect new concept samples, allowing for efficient model updating. Instance weighting adjusts the weights of data samples based on their recency, which helps the learning model adapt to model drift by retraining on weighted samples \cite{drift2}.
\item \textit{Incremental Learning Methods}:  Incremental learning methods continuously update learning models sequentially as new data samples are introduced, ensuring the models stay up-to-date and adapt to evolving data. Hoeffding Trees (HT), Adaptive Online Neural Networks (AONN), and their variants are popular incremental learning methods. HT algorithms, such as Very Fast Decision Tree (VFDT), Concept-adapting VFDT (CVFDT), and Extremely Fast Decision Tree (EFDT), use Hoeffding’s inequality to partially update nodes as new samples arrive, making them suitable for data stream analytics due to their ability to adapt to new samples \cite{drift3}. AONN is another incremental learning method based on neural networks that updates the model when the error rate increases \cite{aonn}. However, it is worth noting that while incremental learning methods offer benefits for adapting to new data, they are not specifically designed to address model drift issues.
\item \textit{Ensemble Learning Methods}: Ensemble learning techniques combine multiple base learners to achieve improved model drift adaptation performance. Block-based ensembles and online ensembles are two main categories of ensemble methods for data stream analytics. Block-based ensembles, such as Streaming Ensemble Algorithm (SEA) \cite{sea}, Accuracy Weighted Ensemble (AWE) \cite{awe},  and Diversity and Transfer-based Ensemble Learning (DTEL) \cite{dtel}, train base learners on discrete data blocks and update the ensemble when new blocks are added to the database. Online ensembles, such as Adaptive Random Forest (ARF) \cite{arf}, Streaming Random Patches (SRP) \cite{srp}, and Performance Weighted Probability Averaging Ensemble (PWPAE) \cite{mypwpae}, integrate multiple incremental learning models to enhance continual learning performance. Ensemble learning models are generally efficient for handling gradual and recurring drifts but may struggle with abrupt drifts and increased computational complexity.
\end{enumerate}

In summary, the detection and adaptation of model drift are essential steps for automated ML model updating to maintain high model performance in dynamic ZTNs. 

\subsection{AutoML-Enabled Autonomous Cybersecurity Mechanisms}\label{S3-sec}
To enhance defense against cyber-attacks, comprehensive cybersecurity systems incorporate four essential steps: anomaly detection, root cause analysis, and system remediation. To enhance defense against cyber-attacks, comprehensive cybersecurity systems incorporate four essential steps: anomaly detection, root cause analysis, system remediation, and intrusion prevention. For their deployment in ZTNs, these procedures could be automated by AutoML techniques to meet the network automation requirements. 
\subsubsection{Automated Anomaly/Intrusion Detection} \label{S3-ids}
Ensuring the reliability and availability of ZTNs and general networks requires the identification and prediction of abnormal behaviors resulting from intentional or malicious actions. Early detection of potential network threats allows for a rapid response, minimizing the risk of intentional harm, service degradation, and monetary loss. AI/ML techniques are recognized as essential enablers for next-generation networks to identify anomalous traffic patterns that could lead to service unavailability or security issues. AI/ML models have demonstrated their ability to develop Intrusion Detection Systems (IDSs) by uncovering hidden patterns in large collections of multi-dimensional, time-varying data \cite{zsmsec3}. In ZTNs and future networks, IDSs should be automatically developed, provisioned, and managed to detect various cyber-attacks.

In the first stage, the ML-based IDSs should enable real-time monitoring of all network data from diverse sources, such as connections, devices, radio networks, current and old core networks, services, transport, and Information technology (IT) operations. Monitoring and collecting network data from a variety of sources enables network security managers to gain a multi-perspective view of the network, which is necessary for identifying anomalies that impact multiple domains or environments \cite{mytnsm}. Supervised ML-based IDSs are capable of detecting intrusions by classifying benign data and known types of attacks, while unsupervised ML-based IDSs can be used to distinguish previously unknown attacks from benign data. AutoML techniques can significantly enhance the performance of both supervised and unsupervised ML-based IDSs by automating the detection of malicious cyber-attacks. Specifically, AutoML can improve network traffic data quality through AutoDP and AutoFE methods, and train optimized IDS models by facilitating automated model selection and HPO techniques \cite{myautoml}. 

The development of IDS systems utilizing ML has captured the attention of the cybersecurity research community, establishing itself as a prime example of the application of ML in enhancing cybersecurity defenses \cite{mythesis}. In their work, Agrafiotis \textit{et al.} \cite{ids2} introduced an ML model combining embeddings with a Fully-Connected network (Embeddings \& FC) specifically designed for identifying malware traffic within 5G networks. This model, when evaluated using the 5G-NIDD dataset tailored for 5G network scenarios, exhibited a notable increase in detection accuracy. Furthermore, Tayfour \textit{et al.} \cite{ids3} explored the use of a Deep Learning (DL) approach through a Long Short-Term Memory (DL-LSTM) architecture aimed at recognizing cyber-attacks targeting IoT and 5G networks, achieving significant accuracy improvements on the CICIDS2017 dataset and showcasing the DL model's capability in accurately identifying network intrusions. Additionally, He \textit{et al.} \cite{ids4} developed an IDS based on a Pyramid Depthwise Separable Convolution neural network (PyDSC-IDS), which, when compared to other DL approaches, delivered superior accuracy in detecting network intrusions with minimal added complexity across various datasets including NSL-KDD, UNSW-NB15, and CICIDS2017, further highlighting the advancements and effectiveness of DL techniques in the realm of intrusion detection systems.

Although the application of AutoML techniques to cybersecurity applications is a relatively new research area, there have been several existing works that have explored using AutoML for the development of autonomous IDSs. The AutoML for Intrusion Detection (AutoML-ID) model proposed by Singh et al. \cite{autoids1} and the Optimized Ensemble IDS (OE-IDS) proposed by Khan et al. \cite{autoids2} represent state-of-the-art approaches that leverage AutoML for intrusion detection. AutoML-ID employs BO for automated model selection and HPO, whereas OE-IDS utilizes the h2o AutoML tool for selecting optimal ML models and developing the final ensemble model. The enhanced accuracy demonstrated by the ML-based IDSs in these studies underscores the potential of AutoML to improve IDS development.

Apart from a comprehensive overview of the network, effective IDSs should provide completely autonomous functionalities capable of continuously and optimally adapting to changes and concept drift. This adaptability can be achieved through AutoML-enabled automated model updating methods, which ensure that the IDSs remain effective over time. By monitoring the complete scope of data and implementing AutoML techniques to develop adaptable and optimized ML-based IDSs, cyber-attacks or network abnormalities are identified more quickly and accurately \cite{myautoml}. 

\subsubsection{Automated Root Cause Analysis} 
Root cause analysis aims to identify the underlying cause of network issues, enabling rapid recovery. By analyzing the entire context and relevant information of the detected abnormal events, this procedure aims to determine the possible causes and appropriate countermeasures/solutions for network anomalies. Root cause analysis often requires a comprehensive correlation analysis across multiple architectural levels, events, environments, and vendors \cite{mytnsm}.

Once an anomaly alert has been triggered by IDSs, it is crucial to determine its root cause. This is necessary to enable self-organizing systems, implement effective mitigation methods, perform network forensics, and assign responsibility. However, with the complexity and diversity of evolving mobile networks, along with the growing number of Key Performance Indicators (KPIs) and data related to end-users, services, and networks, identifying the underlying cause can be challenging. Manual root cause analysis based on expert knowledge is difficult, time-consuming, and labor-intensive. On the other hand, AI/ML models have emerged as an attractive alternative for facilitating self-root cause analysis due to their ability to analyze large amounts of data, identify complex non-linear correlations within the data, and make faster and more accurate decisions compared to manual analysis \cite{zsmsec3}.

In practice, a root cause analysis powered by AutoML facilitates the identification of causative factors for network anomalies, facilitating a prompt recovery and the implementation of effective countermeasures. Through the examination of correlations among various architectural levels, events, environments, and vendors, root cause analysis powered by AutoML can offer comprehensive insights into network problems. This not only enhances network resilience and reliability but also reduces the time and resources required for network diagnostics and recovery. Additionally, the automated model updating procedure in the AutoML process can provide real-time root causes, despite event changes. By leveraging AutoML, networks can autonomously conduct root cause analysis with increased efficiency and precision, thereby mitigating the limitations linked to conventional analysis techniques. 

\subsubsection{Automated System Remediation} 
Network system recovery, which entails the process of restoring a network system to its normal functioning state after being compromised by an attack, is a critical component of network security, facilitating organizations to swiftly respond to security incidents and mitigate the impact of network attacks \cite{recovery1}. This process is closely tied to fault tolerance, ensuring networks maintain functionality amid failures, thus bolstering resilience against disruptions. AutoML-based network monitoring systems pave the way for autonomous remediation by autonomously identifying network anomalies and analyzing their root causes. This information can then be used by the remediation engine to recommend actions for autonomous remediation. Additionally, the remediation operations can be continuously enhanced through fine-tuning in a closed feedback loop.

There are several methods that can recover or remediate a compromised network, including backup and restore, system reimage, network isolation, and blacklisting \cite{mytnsm} \cite{recovery2}. The backup and restore method involves creating regular backups of network systems and restoring them after an attack to recover lost or corrupted data. System reimage involves completely reinstalling the operating system and software on compromised network devices, which proves effective in cases where the attack has caused significant damage to the network system. Network isolation is the process of isolating the compromised network segment from the rest of the network using network segmentation techniques to limit the spread of the attack. This helps prevent further damage and provides a safe environment for recovery efforts. Blacklisting is a common defense method that blocks traffic or requests from detected malicious or suspicious network devices/nodes until further authentication to prevent ongoing attacks. Among AI/ML techniques, Reinforcement Learning (RL) models have been used as a potential autonomous solution to effectively select and recommend actions for fast network recovery \cite{rl1}. Through AutoML techniques, the process of determining the most suitable remediation method or countermeasure can be automated. The automated model updating process can change or revise the remediation method by adapting to different attacks and networking environments.

\section{Adversarial ML Attacks and Defense} \label{S5-AML}

\subsection{Adversarial ML Attacks} 

\begin{figure*}[!t]
\centerline{
\includegraphics[width=11.3cm]{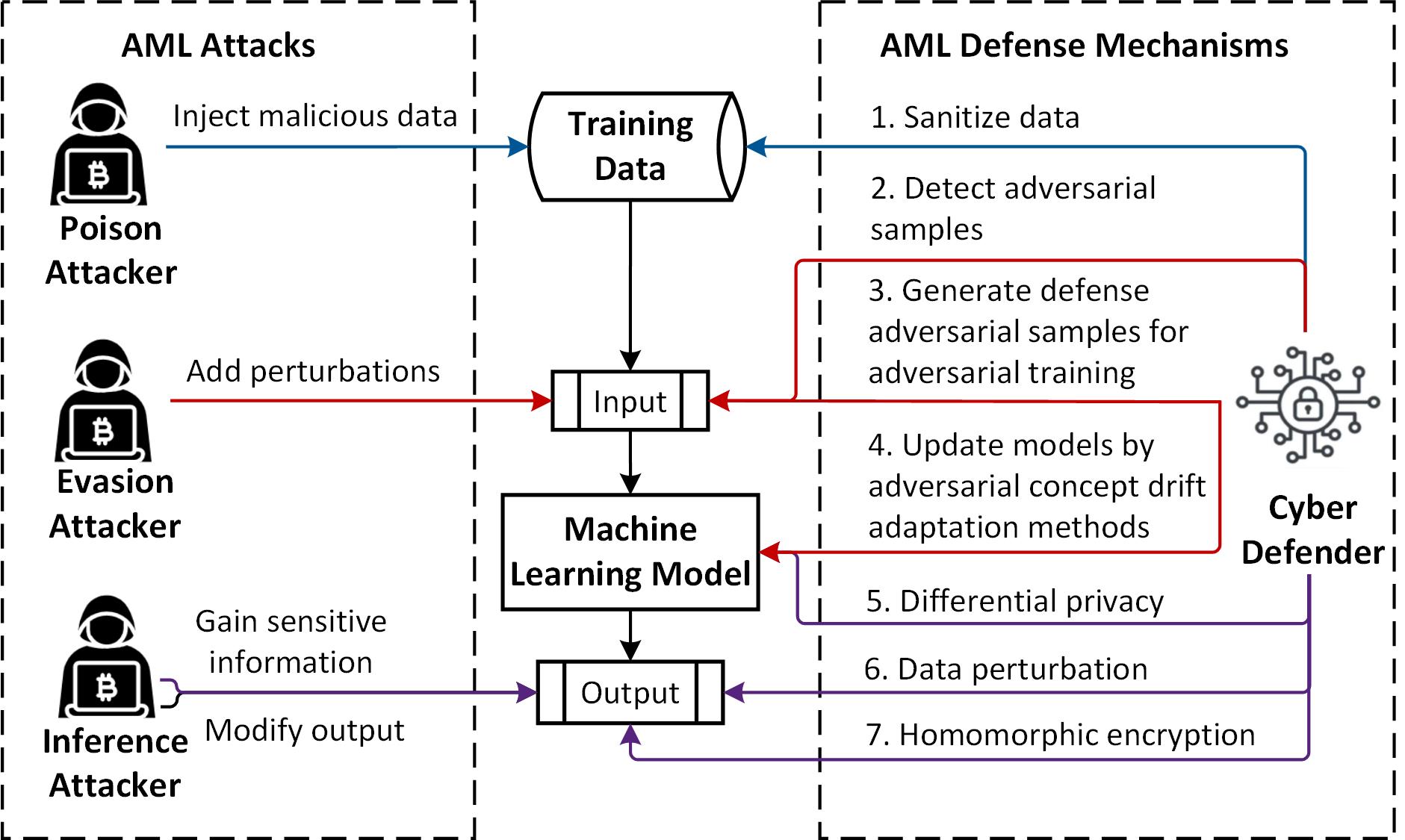}}
\caption{Adversarial ML attacks \& defense for ZTNs.}
\label{aml}
\end{figure*}

The automation of modern networks has experienced significant improvement through the application of AI/ML techniques, resulting in enhanced functionalities such as self-planning, self-optimization, self-healing, and self-protection \cite{zsm6}. However, AI/ML techniques introduce new attack interfaces in ZTNs. The AI/ML models are vulnerable to attacks during the training and testing phases, compromising the integrity, availability, and privacy of the system \cite{aml1}. These targeted cyber threats against AI/ML models are known as AML attacks \cite{aml2}. Although AML attacks are not explicitly designed for ZTNs/6G networks, the increasing demand for network automation and reliance on AI/ML techniques have elevated AML attacks to a significant threat in these networks. Due to network automation requirements and insufficient human supervision, it is challenging for ZTNs to maintain AI/ML model security. Additionally, given the capabilities and extensive uses of AI/ML methods, cyber-criminals can devise complex and advanced cyber-attacks with potentially severe and harmful consequences \cite{zsmsec3}. Therefore, the development of countermeasures to AML attacks is essential to ZTN security. This is particularly important in the AutoML model development process due to minimal human supervision.

In AML attacks, cyber-attackers aim to deceive an AI/ML model by feeding it carefully crafted data intended to trigger inaccurate predictions or classifications. AML attacks can be classified based on the stage at which they occur during the ML process, including data poisoning attacks, evasion attacks, and inference attacks \cite{zsmsec3} \cite{aml3}. Data poisoning attacks occur during the training stage, where the cyber-attacker injects malicious data into the training set to contaminate the model. Evasion attacks occur after the ML model has been trained, where the cyber-attacker manipulates the ML model by modifying its parameters or structure to degrade its performance. Inference attacks occur during the deployment/prediction phase, where the cyber-attacker manipulates the prediction data given to the ML model to cause a specific outcome and potentially gain access to sensitive information. Figure \ref{aml} illustrates these three types of AML attacks and potential defense mechanisms. The specifications for each type of attack are discussed below.

\subsubsection{Poisoning Attacks} 
In poisoning attacks, an attacker manipulates data or the AI/ML model during the training phase to influence the prediction results \cite{aml2}. Attackers can exploit the need for continuous retraining learning models to adapt to new data distributions, thereby creating an opportunity to maliciously manipulate or poison the trained model. In classification problems, poisoning attackers alter the labels of data samples to cause the ML model to misclassify them, while in regression problems, they modify the values of input features to cause the model to produce incorrect outputs \cite{poison1}. Poisoning attacks can be executed using various strategies, such as data injection, data manipulation, and logic corruption \cite{zsmsec3} \cite{aml2}. 

In data injection attacks, the attackers do not have access to the training data and aim to alter the data distribution by introducing crafted malicious samples into the training dataset while keeping the original samples unchanged. This method is effective when the original classifier is trained with limited data and requires additional data for retraining \cite{aml3}.

Data manipulation attacks, on the other hand, assume that attackers have full access to the training data, allowing them to directly contaminate the original data used for training the learning model \cite{zsmsec3}. Contamination involves changing labels (\textit{e.g.}, malicious to benign) or adding small perturbations to the input features. Attackers often target samples or labels with high classification confidence to maximize their impact on the ML model while minimizing the chances of detection \cite{aml2}. 

Logic corruption attacks specifically target the learning algorithm or its underlying logic with the intention of disruption. Logic corruption can be utilized against models that employ distributed learning, such as federated learning, which relies on the training of multiple agents. In such attacks, a malicious agent can manipulate the local model parameters to compromise the global model \cite{zsmsec3}. 

Poisoning attacks can cause the classifier to make erroneous judgments, resulting in reduced accuracy, changes to decision boundaries, and even errors in final test results \cite{6gpri1}. Poisoning attacks pose serious risks to network performance, security, and privacy in 6G networks, which rely significantly on ML for numerous applications and services.

\subsubsection{Evasion Attacks} \label{S5-evasion}
The purpose of evasion attacks is to discover challenging data samples that are most likely to be misclassified to deceive AI/ML models into making incorrect decisions. These attacks, unlike poisoning attacks, do not affect the training process of ML models. Instead, they occur during the testing phase, where the attacker introduces adversarial examples, or minor perturbations, to the input instances  \cite{aml3}. 

Evasion attacks exploit the sensitivity of high-precision ML models to subtle perturbations, rendering ML-enabled machines in autonomous networks vulnerable to being deceived by adversarial examples \cite{6gpri1}. These attacks can result in system misclassification or reduced accuracy, posing privacy risks for ML-based applications in autonomous networks.

Common evasion attacks include Decision Tree Attack (DTA) \cite{dta}, Fast Gradient Sign Method (FGSM) \cite{fgsm}, and Basic Iterative Method (BIM) \cite{bim}. DTA targets decision tree-based ML models (\textit{e.g.}, random forest, XGBoost, and LightGBM) by manipulating input features to mislead the models. This attack demonstrates the need for enhanced protections in tree-based classifiers to maintain their reliability in security-critical applications \cite{dta}. FGSM generates adversarial samples by introducing imperceptible amounts of noise to the input data using gradients derived from the model's loss function \cite{fgsm}. Due to its efficiency, FGSM poses challenges to ML models, particularly DL models. Consequently, it calls for the development of countermeasures to defend against gradient-oriented attacks. BIM is an iterative variant of the FGSM attack in which adversarial perturbations are implemented incrementally over multiple iterations \cite{bim}. By allowing fine-grained control over the perturbation process, BIM can generate more effective adversarial examples. This iterative approach highlights the importance of robust defenses that can withstand iterative gradient-based attacks.

Evasion attacks constitute the primary category of AML attacks, and cyber-attackers employ various techniques to compromise the performance of AI/ML models. Examples of such evasion attacks include the Jacobian-Based Saliency Map Attack (JSMA), Projected Gradient Descent (PGD), Zeroth-Order Optimization (ZOO) attack, and High Confidence Low Uncertainty (HCLU) Attack \cite{art}. 

Defending against evasion attacks is challenging due to various factors, including the high dimensionality and continuous nature of the input space, the difficulty in precisely modeling the decision boundary of the ML models, and the restricted availability of labeled data for training. In addition, the effectiveness of evasion attacks often relies on the particular ML algorithm, its settings, and the quality of the training data.

\subsubsection{Inference Attacks} 
Inference or exploratory attacks refer to AML attacks where the attacker aims to gain knowledge or confidential information about the target ML model, its input data, or its architecture \cite{aml3}. These attacks can be especially harmful in security-critical applications, such as privacy-preserving ML models, where the attacker's goal is to infer sensitive information about individuals or network devices.

Inference attacks include model inversion, model extraction, and membership inference attacks \cite{zsmsec3}.  Model inversion attacks attempt to recover the training data by exploiting the model's outputs. Model extraction attacks seek to reveal the model's architecture and parameters to replicate a nearly identical ML model by observing the model's predictions and/or execution time. Membership inference attacks exploit the model's output to determine whether a data sample was part of the training dataset used for the target ML model. 

Inference/exploratory attacks pose a threat to the privacy and confidentiality of sensitive information contained within the ML models and their training data \cite{6gpri1}. The success of inference attacks often depends on the structure and parameters of the ML model, the quality and quantity of the training data, and the presence of privacy-preserving mechanisms, such as differential privacy. In addition, the efficacy of the attack can be affected by the attacker's knowledge of the ML model and training data.

\subsection{AI/ML Model Security Solutions: AML Defense Mechanisms} 

\begin{table*}[htbp]
\caption{Security Threats in AI/ML Models.}
\setlength\extrarowheight{1pt}
\begin{tabularx}{\textwidth}{|>{\centering\arraybackslash}p{1.2cm}|>{\centering\arraybackslash}p{3.2cm}|>{\centering\arraybackslash}p{6.5cm}|>{\centering\arraybackslash}X|}
\hline
\textbf{Attack Category} & \textbf{Specific Attack Examples} & \textbf{Description} & \textbf{Countermeasures} \\ \hline
\multirow{3}{=}{\centering Poisoning Attacks} & Data Injection & \multirow{3}{=}{Occur during the training phase, manipulate data or learning algorithms to influence prediction results.} & \multirow{3}{=}{Input validation, data sanitization, anomaly detection, etc.} \\ \cline{2-2}
& Data Manipulation & & \\ \cline{2-2}
& Logic Corruption & & \\ \hline
\multirow{3}{=}{\centering Evasion Attacks} & Decision Tree Attack & \multirow{3}{=}{Occur after the training phase, discover challenging data samples likely to be misclassified to deceive AI/ML models.} & \multirow{3}{=}{Adversarial training, adversarial sample detection, ensemble methods, adversarial concept drift adaptation, etc.} \\ \cline{2-2}
& Fast Gradient Sign Method & & \\ \cline{2-2}
& Basic Iterative Method & & \\ \hline
\multirow{3}{=}{\centering Inference Attacks} & Model Inversion & \rule[-0pt]{0pt}{9pt}\multirow{1}{=}{Occur during the deployment/prediction phase, aiming to gain knowledge or confidential information about the target ML model, its training data, or its architecture.} & \multirow{1}{=}{Differential Privacy (DP), data perturbation, homomorphic encryption, etc.} \\ \cline{2-2}
& Model Extraction & & \\ \cline{2-2}
& Membership Inference & & \\ \hline
\end{tabularx}
\label{5-aml}
\end{table*}

AML defense is an emerging research field focused on enhancing the resilience of ML approaches against adversarial attacks \cite{aml2}. Its primary objective is to assess the vulnerability of ML algorithms to adversarial attacks and develop effective responses that promote more robust learning. Several defense mechanisms have been proposed to address adversarial attacks, such as input validation, adversarial training, adversarial sample detection, defense Generative Adversarial Networks (GANs), concept drift adaptation, differential privacy, and homomorphic encryption \cite{zsmsec1}, as illustrated in Fig. \ref{aml}. 
\subsubsection{Countermeasures Against Poisoning Attacks}
There are several potential countermeasures/solutions to defend against poisoning attacks. Input validation involves the sanitization of training data by eliminating malicious or unusual samples before integrating them into the ML model \cite{zsmsec3}. Data sanitization and anomaly detection methods are also used as countermeasures to remove suspicious/malicious samples from the training data \cite{poison1}. 
\subsubsection{Countermeasures Against Evasion Attacks}
Evasion attacks can be defeated by adversarial training, adversarial sample detection, ensemble methods, and adversarial concept drift adaptation techniques \cite{zsmsec3} \cite{eva1}. Adversarial training can train/retain AI/ML models on an augmented dataset with adversarial samples to improve model performance. Adversarial samples can be generated by defense GANs. In adversarial sample detection countermeasures, the adversarial samples generated by AML attacks are detected and removed from the input data. Ensemble learning methods can improve the robustness of AI/ML models and reduce the impact of polluted models by integrating multiple base learners. Adversarial concept drift adaptation techniques can update or retrain AI/ML models to confront AML attacks once a data distribution change or a model performance degradation is detected.  
\subsubsection{Countermeasures Against Inference Attacks}
To mitigate the impact of inference attacks, researchers have proposed several approaches, including Differential Privacy (DP), data perturbation, and homomorphic encryption \cite{inf1}. In DP methods, random noise is injected into the model's outputs to prevent the attacker from gaining knowledge about the training data or individual samples. Data perturbation aims to perturb or transform the training data to make it difficult for the attacker to infer sensitive information. Homomorphic encryption enables model training with encrypted data, ensuring the confidentiality of data. However, this approach introduces computational complexities that need to be carefully managed \cite{zsmsec3}.

\subsection{Summary}
The common types of AML attacks and their potential countermeasures are summarized in Table \ref{5-aml}. Although there are many countermeasures to general AML attacks, cyber-attackers have been continually evolving their tactics to evade these defenses and breach ML models. Therefore, the development of effective defense mechanisms for AML attacks in ZTNs is critical to ensuring the security and robustness of ML models integrated into ZTNs/6G networks.

\section{Case Study} \label{S6}
With the introduction of ZTNs, AutoML techniques, and AML attacks, this section presents two case studies to illustrate the capabilities and benefits of applying AutoML techniques to ZTN security and depict the typical AML attack and defense process.
The first case study presents a detailed AutoML pipeline designed for automated intrusion detection within ZTNs. In the second case study, we simulate three common adversarial attacks and devise basic defense mechanisms to safeguard the IDS model obtained from the first study against these attacks. The experimental results analysis for each case study are presented in Sections \ref{S6-U1} and \ref{S6-U2}.

\subsection{Description of Two Use Cases}
With the rise of 5G networks, the increase in connected devices and data traffic has broadened the attack surface, making networks more vulnerable to cyber threats. The evolution towards ZTN and future networks necessitates communication systems that are not only more secure but also more autonomous and efficient. IDSs, as discussed in Section \ref{S3-ids}, play a pivotal role in monitoring and detecting security threats and malicious activities within modern networks \cite{zsmsec3}.

However, ZTN security solutions require high levels of automation and minimal human intervention \cite{zsmsec1}. Therefore, the development of automated IDSs becomes essential to achieving optimal network management in ZTNs. As discussed in Section \ref{S5-AutoML}, AutoML is an advanced ML technology that automates the design and implementation of ML models. This improves traditional ML procedures, making them more autonomous and optimized. By leveraging AutoML techniques, it is possible to develop efficient, accurate, and adaptable IDSs that overcome the performance and human supervision limitations of traditional IDSs \cite{automl6}. 

Therefore, the first case study aims to provide a comprehensive analysis of the development and implementation of an AutoML-based IDS for ZTNs. By leveraging ML algorithms and optimization methods outlined in Sections \ref{S5-ML} and \ref{S5-AutoML}, the case study aims to establish optimal classification models capable of distinguishing malicious attacks from benign events. Additionally, the online learning methods introduced in Section \ref{AMU} are used in the online AutoML-based IDS development for realizing the automated model updating functionalities of AutoML in dynamic ZTN environments.

The second case study focuses on the cyber-defense exercise of AML attacks targeting the AutoML-based IDS derived from the first case study. As many ZTN services and functionalities rely on AI/ML models, they are becoming increasingly vulnerable to AML attacks. AML attacks exploit the weaknesses and vulnerabilities of ML models by generating adversarial inputs that can deceive or manipulate the models into making incorrect predictions \cite{aml2}. In ZTNs, AML attacks pose a significant threat to overall network security and reliability. This case study aims to demonstrate the detrimental impact that AML attacks can have on ML models in ZTNs, and presents basic defense strategies to mitigate these attacks, thereby ensuring the accuracy of the ML-based IDS. In this case study, three common types of adversarial attacks discussed in Section \ref{S5-evasion} (\textit{i.e.}, DTA \cite{dta}, FGSM \cite{fgsm}, and BIM \cite{bim}) are used to generate adversarial samples to probe the vulnerability of the IDS. Subsequently, the basic defense mechanisms, including adversarial sample detection and filtering, are devised to safeguard the AutoML-based IDS for ZTNs against AML attacks.

Two public benchmark network security datasets are utilized in these case studies to evaluate the proposed AutoML-based IDS and the AML attack \& defense models: the Canadian Institute for Cybersecurity Intrusion Detection System 2017 (CICIDS2017) dataset \cite{cic} and the 5th Generation Network Intrusion Detection Dataset (5G-NIDD) \cite{5gnidd}. The CICIDS2017 dataset is a state-of-the-art intrusion detection dataset with the latest network threats. The CICIDS2017 dataset is close to real-world network data since it has a large amount of network traffic data, a variety of network features (80), various types of attacks (14), and highly imbalanced classes. The primary types of attacks in the CICIDS2017 dataset include DoS/DDoS, botnets, brute force, port scan, and web attacks. 

The 5G-NIDD dataset, created in December 2022, is a fully labeled resource constructed on a functional 5G test network for researchers and practitioners evaluating AI/ML solutions in the context of 5G/6G security \cite{5gnidd}. 5G-NIDD encompasses data extracted from a 5G testbed connected to the 5G Test Network (5GTN) at the University of Oulu, Finland. The dataset is derived from two base stations, each featuring an attacker node and multiple benign 5G users. The attacker nodes target a server deployed within the 5GTN MEC environment. The attack scenarios captured in the dataset primarily include DoS attacks and port scans. By employing these two up-to-date datasets that closely mirror real-world network traffic, the case study results become highly relevant to intrusion detection tasks and challenges in ZTNs. The case studies are conducted using a reduced CICIDS2017 dataset with 28,307 samples and a reduced 5G-NIDD dataset with 12,159 instances for the purpose of this work. 

The AutoML-based IDSs consist of offline learning and online adaptive learning functionalities. To evaluate the offline ML/AutoML models in the case studies, 5-fold cross-validation is used in the experiments, as it is a commonly-used split that can help maintain a balance between bias and variance and reduce over-fitting \cite{split}. In the experiments evaluating the long-term online learning performance of drift-adaptive models on network data in dynamic networking environments, prequential evaluation, or called test-and-train evaluation, is utilized. In prequential validation, each new data sample undergoes an initial real-time test by the learners. Then, the learning model incorporates the sample for potential updates \cite{mypwpae}. Prequential validation is acknowledged as the standard approach for assessing online learning processes.

To gauge the effectiveness of the models, a combination of four classification performance metrics is employed: accuracy, precision, recall, and F1-scores. These four classification metrics are often used in ML research as they provide a well-rounded view of ML model performance \cite{metric}. Due to the inherent imbalance in intrusion detection datasets, relying solely on individual performance metrics, such as accuracy, precision, or recall, can lead to a partial understanding of model performance. Therefore, we considered accuracy, precision, recall, and F1-scores collectively for a comprehensive comparison of ML models to avoid skewed evaluation results. The F1-score, which calculates the harmonic mean of recall and precision scores, provides a balanced view of anomaly detection outcomes while reducing bias. By including both false negatives (assessed by recall) and false positives (measured by precision) in its calculation, the F1-score offers a comprehensive performance metric for evaluating the proposed AutoML pipeline. 

Furthermore, to cater to the processing time and efficiency demands of network systems, the learning time of each model is analyzed to compare the learning efficiency of the ML models. The ideal ML model for network data analytics should strike a balance between effectiveness and efficiency, ensuring optimal performance while minimizing computational overhead.

\subsection{Use Case 1: AutoML-based Automated IDS} \label{S6-U1}

\begin{table*} [htbp]
\caption{The specifications of the proposed AutoML pipeline.}
\setlength\extrarowheight{1pt}
\centering
\begin{tabular}{|>{\centering\arraybackslash}p{1.3cm}|>{\centering\arraybackslash}p{2cm}|>{\centering\arraybackslash}p{2.5cm}|p{10cm}|}
\hline
\textbf{Category}                         & \textbf{Procedure}                        & \textbf{Method}                  & {\textbf{Aim/Operation}}                                                                                                                                                                   \\ 
\hline
\multirow{6}{*}{AutoDP}                   & Encoding                                  & Label Encoding                   & Convert categorical features into numerical features to enhance the readability of the data for ML models.                                                                           \\ 
\cline{2-4}
                                          & Imputation                                & Median Imputation                  & Identify and impute missing values with the median values to improve data quality.                                                                                                                               \\ 
\cline{2-4}
                                          & Normalization                             & Z-Score or Min-Max Normalization & Automatically select a method to normalize features to a similar scale to enhance data quality.                                                                                                             \\ 
\cline{2-4}
                                          & Balancing                                 & ADASYN                            & Generate synthetic minority class samples to balance data and enhance data quality.                                                                                                      \\ 
\hline
\multirow{2}{*}{AutoFE}                   & \multirow{2}{*}{Feature Selection}        & RFE                               & Select important features and remove irrelevant features to improve model efficiency                                                                                                                                 \\ 
\cline{3-4}
                                          &                                           & Pearson Correlation              & Remove redundant features to improve model efficiency and accuracy                                                                                                                     \\ 
\hline
\multirow{5}{*}{\shortstack{Automate\\\\ Model\\\\ Learning}} & \multirow{4}{*}{\shortstack{Offline Model\\\\ Selection}}          & KNN                               & \multirow{4}{9cm}{Select the best-performing model among four alternative offline ML models by evaluating their learning performance.}                                                               \\ 
\cline{3-3}
                                          &                                           & MLP                              &                                                                                                                                                                                          \\ 
\cline{3-3}
                                          &                                           & RF                               &                                                                                                                                                                                          \\ 
\cline{3-3}
                                          &                                           & LightGBM                         &                                                                                                                                                                                          \\ 
\cline{2-4}
                                          & HPO               & PSO                           & Tune the hyperparameters of the offline ML models to optimize them.                                                                                                         \\ 
\hline
\multirow{5}{*}{\shortstack{Automated\\\\ Model\\\\ Updating}} & \multirow{4}{*}{\shortstack{Online Model\\\\ Selection}} & HT                                & \multirow{4}{9cm}{Select the best-performing model among four online drift-adaptive models to adapt to dynamic network data streams with concept drift issues}  \\ 
\cline{3-3}
                                          &                                           & EFDT                             &                                                                                                                                                                                          \\ 
\cline{3-3}
                                          &                                           & ARF                              &                                                                                                                                                                                          \\ 
\cline{3-3}
                                          &                                           & SRP                              &                                                                                                                                                                                          \\ 
\cline{2-4}
                                          & HPO               & PSO                           & Tune the hyperparameters of the online ML models to optimize them.                                                                                                         \\ 
\hline
\end{tabular}
\label{6-automl}%
\end{table*}

\subsubsection{Experimental Setup} 
The first case study utilizes a comprehensive AutoML pipeline to achieve automated intrusion detection for both static and dynamic ZTN environments. 

The experiments on AutoML-based IDS development were run on a machine with an i7-8700 processor and 16 GB of memory, representing a server machine in ZTNs for deploying autonomous IDSs. The functions and methods used in this case study are implemented by extending the following Python libraries: Sklearn \cite{sklearn} and LightGBM \cite{lightgbm} for static ML model development, Optunity \cite{Optunity} for ML model optimization and automation, and River \cite{river} for online ML model development and automated model updating.

The proposed AutoML-based IDSs enable the automation of all necessary procedures in ML/data analytics, including automated data pre-processing, automated feature engineering, automated model selection, hyperparameter optimization, and automated model updating (drift adaptation). 
Overall, the procedures, methods, and aims/operations of each stage of the proposed AutoML pipeline are summarized in Table \ref{6-automl}.

As illustrated in Table \ref{6-automl}, the AutoML pipeline begins with AutoDP, which involves these steps:
\begin{enumerate}[label=\roman*)]
\item \textit{Encoding}: The pipeline automatically detects categorical features and transforms them into numerical features using label encoding. This enhances the data's readability for ML models.
\item \textit{Imputation}: It automatically spots missing values in the datasets and fills them with the median values using median imputation. This step enhances data quality.
\item \textit{Normalization}: The pipeline chooses an appropriate normalization method (z-score or min-max normalization) to bring features to a similar scale. This enhances data quality.
\item \textit{Balancing}: The proposed system automatically identifies whether the dataset is imbalanced. If an imbalance is identified, synthetic minority class samples will be generated using the ADASYN method \cite{adasyn} to balance the data and improve data quality.
\end{enumerate}

Next, the pipeline enters the AutoFE stage. In this stage, the RFE method is used to select vital features and discard irrelevant or noisy features. RFE is a wrapper feature selection method that recursively generates and evaluates feature subsets to remove unimportant features until a defined number of features is selected \cite{rfe}. To automate this step, PSO is used to optimize the number of selected features as a hyperparameter.

After AutoDP and AutoFE, the offline AutoML pipeline enters the automated model learning stage. In the first step of this stage, the pipeline automatically selects learning models from a set of four representative ML models: K-Nearest Neighbors (KNN), Multilayer Perceptron (MLP), Random Forest (RF), and LightGBM. KNN serves a basic and representative low-complexity ML algorithm, while MLP represents a basic DL model commonly used in various applications. RF and LightGBM are two representative ensemble ML models known for effectively handling non-linear and complex network data. Their widespread usage in data analytics applications attests to their high generalizability. After assessing the performance of each learning model in terms of accuracy and F1-scores, both the top-performing model and the second best-performing model with default hyperparameters are selected for further evaluation through HPO. This selection strategy, incorporating both the top and second best-performing models, mitigates the risk of overlooking the true optimal model and increases the likelihood of identifying the most suitable one.

Once the top two learning models are identified, the HPO technique is applied to fine-tune their hyperparameters, yielding two optimized models. From these, the final optimal model is chosen. PSO is selected as the HPO method due to its superior performance in handling large hyperparameter spaces, low complexity, and generalization ability \cite{myhpo}.

On the other hand, in dynamic networking environments, an additional step, automated model updating, is required to address model drift issues and maintain the continuous reliability of ML models. Correspondingly, online learning methods are required to replace static ML models in this case. In the automated model updating or online learning process, the online AutoML pipeline automatically selects online learning models from four representative online models: HT \cite{drift3}, KNN-ADWIN \cite{knnadwin}, ARF \cite{arf}, and SRP \cite{srp}, as introduced in Section \ref{AMU}. Similar to offline learning, the online AutoML pipeline will select the top two best-performing models for further HPO by PSO to obtain the optimized online learning model for intrusion detection in dynamic ZTNs.

\subsubsection{Experimental Results and Analysis}
The effectiveness of the AutoML framework was assessed by comparing the accuracy, precision, recall, F1-score, and model learning time of ML models developed with and without using AutoML techniques. The comparative performance of the ML/AutoML models on the CICIDS2017 and the 5G-NIDD datasets is demonstrated in Table \ref{result1}.

\begin{table*}[htbp]
\centering%
\setlength\extrarowheight{1pt}
\caption{The experimental results of the proposed offline AutoML-based IDS on the CICIDS2017 and 5G-NIDD datasets.}
\scalebox{0.85}{
\begin{tabular}{|>{\centering\arraybackslash}p{1.5cm}|>{\centering\arraybackslash}p{1.6cm}|>{\centering\arraybackslash}p{1.1cm}|>{\centering\arraybackslash}p{1.1cm}|>{\centering\arraybackslash}p{1cm}|>{\centering\arraybackslash}p{1cm}|>{\centering\arraybackslash}p{1.5cm}|>{\centering\arraybackslash}p{1.1cm}|>{\centering\arraybackslash}p{1.1cm}|>{\centering\arraybackslash}p{1cm}|>{\centering\arraybackslash}p{1cm}|>{\centering\arraybackslash}p{1.5cm}|}

\hline
\multirow{3}{*}{\textbf{\shortstack{AutoML\\\\ Procedures}}} & \multirow{3}{*}{\textbf{\shortstack{Offline\\ Learning\\ Algorithm}}} & \multicolumn{5}{c|}{\textbf{CICIDS2017 Dataset}}                                                                                                                                                                  & \multicolumn{5}{c|}{\textbf{5G-NIDD Dataset}}                                                                                                                                                                     \\ \cline{3-12} 
                                            &                                              & {\textbf{Accuracy (\%)}} & {\textbf{Precision (\%)}} & {\textbf{Recall (\%)}} & {\textbf{F1 (\%)}} & \textbf{Model Learning Time (s)} & {\textbf{Accuracy (\%)}} & {\textbf{Precision (\%)}} & {\textbf{Recall (\%)}} & {\textbf{F1 (\%)}} & \textbf{Model Learning Time (s)} \\ \hline
\multirow{4}{*}{No}                         & KNN                                          & {97.238}                 & {92.081}                  & {93.782}               & {92.923}           & 3.9                              & {99.301}                 & {99.198}                  & {99.664}               & {99.430}           & 0.6                              \\ \cline{2-12} 
                                            & MLP                                          & {88.536}                 & {94.277}                  & {43.701}               & {58.830}           & 15.1                             & {98.026}                 & {99.587}                  & {97.179}               & {98.368}           & 7.6                              \\ \cline{2-12} 
 & \textbf{RF}                                  & {\textbf{99.703}}        & {\textbf{99.577}}         & {\textbf{98.830}}      & {\textbf{99.248}}  & \textbf{3.0}                     & {\textbf{99.670}}        & {\textbf{99.933}}         & {\textbf{99.530}}& {\textbf{99.731}}  & \textbf{3.3}                     \\ \cline{2-12} 
                                            & \textbf{LightGBM}                            & {\textbf{99.816}}        & {\textbf{99.543}}         & {\textbf{99.506}}      & {\textbf{99.525}}  & \textbf{0.3}                     & {\textbf{99.630}}        & {\textbf{99.798}}         & {\textbf{99.597}}      & {\textbf{99.697}}  & \textbf{0.8}                     \\ \hline
\multirow{4}{*}{\shortstack{AutoDP \& \\\\  AutoFE}}         & KNN                                          & {97.058}                 & {92.024}                  & {92.831}               & {92.423}           & 0.5                              & {99.630}                 & {99.465}                  & {99.933}               & {99.698}           & 0.6                              \\ \cline{2-12} 
                                            & MLP                                          & {85.968}                 & {92.069}                  & {26.563}               & {44.831}           & 16.1                             & {98.808}                 & {99.863}                  & {98.187}               & {99.018}           & 6.3                              \\ \cline{2-12} 
& \textbf{RF}                                  & {\textbf{99.735}}        & {\textbf{99.465}}         & {\textbf{99.200}}      & {\textbf{99.332}}  & \textbf{2.8}                     & {\textbf{99.794}}        & {\textbf{99.732}}         & {\textbf{99.933}}      & {\textbf{99.832}}  & \textbf{0.6}                     \\ \cline{2-12} 
                                            & \textbf{LightGBM}                            & {\textbf{99.770}}        & {\textbf{99.466}}         & {\textbf{99.378}}      & {\textbf{99.422}}  & \textbf{0.2}                     & {\textbf{99.794}}        & {\textbf{99.866}}         & {\textbf{99.799}}      & {\textbf{99.832}}  & \textbf{0.2}                     \\ \hline
\multirow{2}{*}{All}                        & RF                                           & {99.735}                 & {99.465}                  & {99.200}               & {99.332}           & 3.6                              & {99.836}                 & {99.866}                  & {99.866}               & {99.866}           & 1.3                              \\ \cline{2-12} 
& \textbf{LightGBM}                            & {\textbf{99.823}}        & {\textbf{99.468}}         & {\textbf{99.644}}      & {\textbf{99.556}}  & \textbf{0.2}                     & {\textbf{99.877}}        & {\textbf{99.799}}         & {\textbf{100.0}}       & {\textbf{99.899}}  & \textbf{0.7}                     \\ \hline
\end{tabular}
}
\label{result1}
\end{table*}



\begin{table} [tp]
\caption{The HPO configuration of the best-performing learning models.}
\setlength\extrarowheight{1pt}
\centering
\scalebox{0.85}{
\begin{tabular}{|>{\centering\arraybackslash}p{1.15cm}|>{\centering\arraybackslash}p{2.1cm}|>{\centering\arraybackslash}p{1.7cm}|>{\centering\arraybackslash}p{1.6cm}|>{\centering\arraybackslash}p{1.3cm}|}
\hline
\textbf{Model}            & \textbf{Hyperparameter Name} & \textbf{Configuration Space} & \textbf{Optimal Value on CICIDS2017} & \textbf{Optimal Value on 5G-NIDD}  \\ 
\hline

\multirow{5}{*}{RF}       & n\_estimators                & {[}50,500]                   & 460                                  & 200                                \\ 
\cline{2-5}
                          & max\_depth                   & {[}5,50]                     & 25                                   & 38                                 \\ 
\cline{2-5}
                          & min\_samples\_split          & {[}2,11]                     & 4                                    & 5                                  \\ 
\cline{2-5}
                          & min\_samples\_leaf           & {[}1,11]                     & 1                                    & 2                                  \\ 
\cline{2-5}
                          & criterion                    & {[}’gini’, ’entropy’]        & ’entropy’                            & ’entropy’                          \\ 
\hline
\multirow{5}{*}{LightGBM} & n\_estimators                & {[}50,500]                   & 300                                  & 340                                \\ 
\cline{2-5}
                          & max\_depth                   & {[}5,50]                     & 42                                   & 22                                 \\ 
\cline{2-5}
                          & learning\_rate               & (0, 1)                       & 0.884                                & 0.101                              \\ 
\cline{2-5}
                          & num\_leaves                  & {[}100,2000]                 & 1400                                 & 1300                               \\ 
\cline{2-5}
                          & min\_child\_samples          & {[}10,50]                    & 30                                   & 45                                 \\ 
\hline
\multirow{2}{*}{ARF}      & n\_models                    & {[}3, 10]                    & 6&  5\\ 
\cline{2-5}
                          & drift\_detector              & {[}‘ADWIN’, ‘DDM’]           & ‘DDM’                                & ‘DDM’                              \\ 
\hline
\multirow{2}{*}{SRP}      & n\_models                    & {[}3, 10]                    & 4& 5                                 \\ 
\cline{2-5}
                          & drift\_detector              & {[}‘ADWIN’, ‘DDM’]           & ‘DDM’                                & ‘DDM’                              \\
                          \hline
\end{tabular}
}
\label{hpo}%
\end{table}

Table \ref{result1} demonstrates the effectiveness of the proposed AutoML pipeline by presenting three sets of results. The first set of results, "AutoML Procedures: No", demonstrates the performance of four original ML algorithms with default hyperparameter configurations (without AutoML) as a baseline. The second set of results, "AutoML Procedures: AutoDP \& AutoFE", displays the performance of ML algorithms after applying the proposed AutoDP and AutoFE procedures, emphasizing the impact of data quality enhancement through AutoML. Finally, the third set, "AutoML Procedures: All", exhibits the performance of the complete AutoML pipeline, which includes AutoDP, AutoFE, automated model selection of the top-2 ML algorithms, and HPO procedures.

The AutoML pipeline begins with the execution of AutoDP and AutoFE, followed by the automated selection of the two best-performing learning models based on their F1-scores in the second set of results. Subsequently, the hyperparameters of the chosen models are optimized to obtain an ultimate optimal model with the highest F1-score, as illustrated in the third set of results. The top-performing configurations in each experimental set are highlighted in bold in Table \ref{result1}.

During the automated model learning process outlined in Table \ref{6-automl}, the two best-performing models are selected as candidate models, and their hyperparameters are tuned using PSO to obtain the ultimate optimal learning model. As demonstrated in Table \ref{result1}, the two most effective ML algorithms for both datasets are RF and LightGBM; hence, their hyperparameters are optimized. The search space and optimal values for the hyperparameters of these two learning algorithms are provided in Table \ref{hpo}. Continuous hyperparameters are assigned a search range, while categorical hyperparameters are given all potential values/options. 

The left part of Table \ref{result1} summarizes the results of the proposed AutoML pipeline on the CICIDS2017 dataset. For the original ML models without any AutoML procedures, four models (KNN, MLP, RF, and LightGBM) exhibit large performance differences. The original RF and LightGBM models significantly outperform the other two ML models. After implementing the proposed AutoDP and AutoFE procedures, RF and LightGBM models remain the top-performing models, surpassing the other models by over 6\% in terms of the F1-score. The data quality improvement caused by AutoDP and AutoFE results in an increase in F1-scores and a decrease in learning time for both RF and LightGBM models. The complete AutoML pipeline further boosts the performance of these learning models, with the optimized LightGBM model achieving the highest F1-score of 99.823\%.

Similarly, the right part of Table \ref{result1} reveals that RF and LightGBM models outperform the other two compared ML models on the 5G-NIDD dataset. Implementing AutoDP and AutoFE results in slight F1-score improvements and reduced learning time for each ML model. After applying HPO to the two best-performing models, RF and LightGBM, the optimized LightGBM model attains the highest F1-score of 99.877\% and is chosen as the final optimal model for the 5G-NIDD dataset.

In conclusion, the offline AutoML pipeline generates a superior IDS model with F1-score improvements of 0.134\% and 0.067\%, and learning time reductions of 50\% and 12.5\% compared to the best-performing learning model without AutoML on the CICIDS2017 and 5G-NIDD datasets, respectively. This highlights the effectiveness of the AutoML pipeline in improving the quality of the IDS by effectively automating the ML process.

\begin{table*}[htbp]
\centering%
\setlength\extrarowheight{1pt}
\caption{The experimental results of the proposed online AutoML-based IDS on the CICIDS2017 and 5G-NIDD datasets.}
\scalebox{0.85}{
\begin{tabular}{|>{\centering\arraybackslash}p{1.5cm}|>{\centering\arraybackslash}p{1.6cm}|>{\centering\arraybackslash}p{1.1cm}|>{\centering\arraybackslash}p{1.1cm}|>{\centering\arraybackslash}p{1cm}|>{\centering\arraybackslash}p{1cm}|>{\centering\arraybackslash}p{1.5cm}|>{\centering\arraybackslash}p{1.1cm}|>{\centering\arraybackslash}p{1.1cm}|>{\centering\arraybackslash}p{1cm}|>{\centering\arraybackslash}p{1cm}|>{\centering\arraybackslash}p{1.5cm}|}

\hline
\multirow{3}{*}{\textbf{\shortstack{AutoML\\\\ Procedures}}} & \multirow{4}{*}{\textbf{\shortstack{Online\\ Learning\\ Algorithm}}} & \multicolumn{5}{c|}{\textbf{CICIDS2017 Dataset}}                                                                                                                                                                  & \multicolumn{5}{c|}{\textbf{5G-NIDD Dataset}}                                                                                                                                                                     \\ \cline{3-12} 
                                            &                                              & {\textbf{Accuracy (\%)}} & {\textbf{Precision (\%)}} & {\textbf{Recall (\%)}} & {\textbf{F1 (\%)}} & \textbf{Model Learning Time (s)} & {\textbf{Accuracy (\%)}} & {\textbf{Precision (\%)}} & {\textbf{Recall (\%)}} & {\textbf{F1 (\%)}} & \textbf{Model Learning Time (s)} \\ \hline
\multirow{4}{*}{No}                         & KNN-ADWIN \cite{knnadwin}                                          & {96.913}& {90.112}& {94.782}& {92.388}& 12.8& {98.729}& {98.469}& {99.453}& {98.958}& 6.0\\ \cline{2-12} 
                                            & HT \cite{drift3}                                         & {90.839}& {73.092}& {84.904}& {78.557}& 6.0& {96.827}& {97.283}& {97.496}& {97.389}& 4.7\\ \cline{2-12} 
 & \textbf{ARF}  \cite{arf}                                & {\textbf{98.451}}& {\textbf{96.215}}& {\textbf{95.937}}& {\textbf{96.073}}& \textbf{22.9}& {\textbf{98.696}}& {\textbf{98.772}}& {\textbf{99.083}}& {\textbf{99.927}}& \textbf{10.6}\\ \cline{2-12} 
                                            & \textbf{SRP} \cite{srp}                            & {\textbf{98.547}}& {\textbf{95.968}}& {\textbf{96.714}}& {\textbf{96.340}}& \textbf{80.0}& {\textbf{99.261}}& {\textbf{99.438}}& {\textbf{99.343}}& {\textbf{99.391}}& \textbf{45.4}\\ \hline
\multirow{4}{*}{\shortstack{AutoDP \& \\\\  AutoFE}}         & KNN-ADWIN                                          & {98.08}& {93.768}& {96.714}& {95.218}& 11.9& {98.771}& {98.404}& {99.589}& {98.993}& 4.9\\ \cline{2-12} 
                                            & HT                                          & {93.173}& {80.083}& {87.125}& {83.456}& 5.4& {97.068}& {98.480}& {96.661}& {97.562}& 2.5\\ \cline{2-12} 
& \textbf{ARF}                                  & {\textbf{98.890}}& {\textbf{97.683}}& {\textbf{96.678}}& {\textbf{97.178}}& \textbf{24.4}& {\textbf{98.987}}& {\textbf{99.044}}& {\textbf{99.2883}}& {\textbf{99.166}}& \textbf{9.4}\\ \cline{2-12} 
                                            & \textbf{SRP}                            & {\textbf{99.008}}& {\textbf{98.452}}& {\textbf{96.497}}& {\textbf{97.465}}& \textbf{58.4}& {\textbf{99.269}}& {\textbf{99.357}}& {\textbf{99.439}}& {\textbf{99.398}}& \textbf{24.0}\\ \hline
\multirow{2}{*}{All}                        & ARF                                           & {\textbf{99.090}}& {\textbf{98.512}}& {\textbf{96.858}}& {\textbf{97.678}}& \textbf{30.8}& {99.227}& {99.411}& {99.316}& {99.363}& 19.1\\ \cline{2-12} 
& \textbf{SRP}                            & {\textbf{99.254}}& {\textbf{99.133}}& {\textbf{99.075}}& {\textbf{99.093}}& \textbf{68.2}& {\textbf{99.360}}        & {\textbf{99.548}}         & {\textbf{99.398}}      & {\textbf{99.473}}  & \textbf{33.9}                     \\ \hline
\end{tabular}
}
\label{result11}
\end{table*}

\begin{figure}[!t]
\centerline{
\includegraphics[width=\columnwidth]{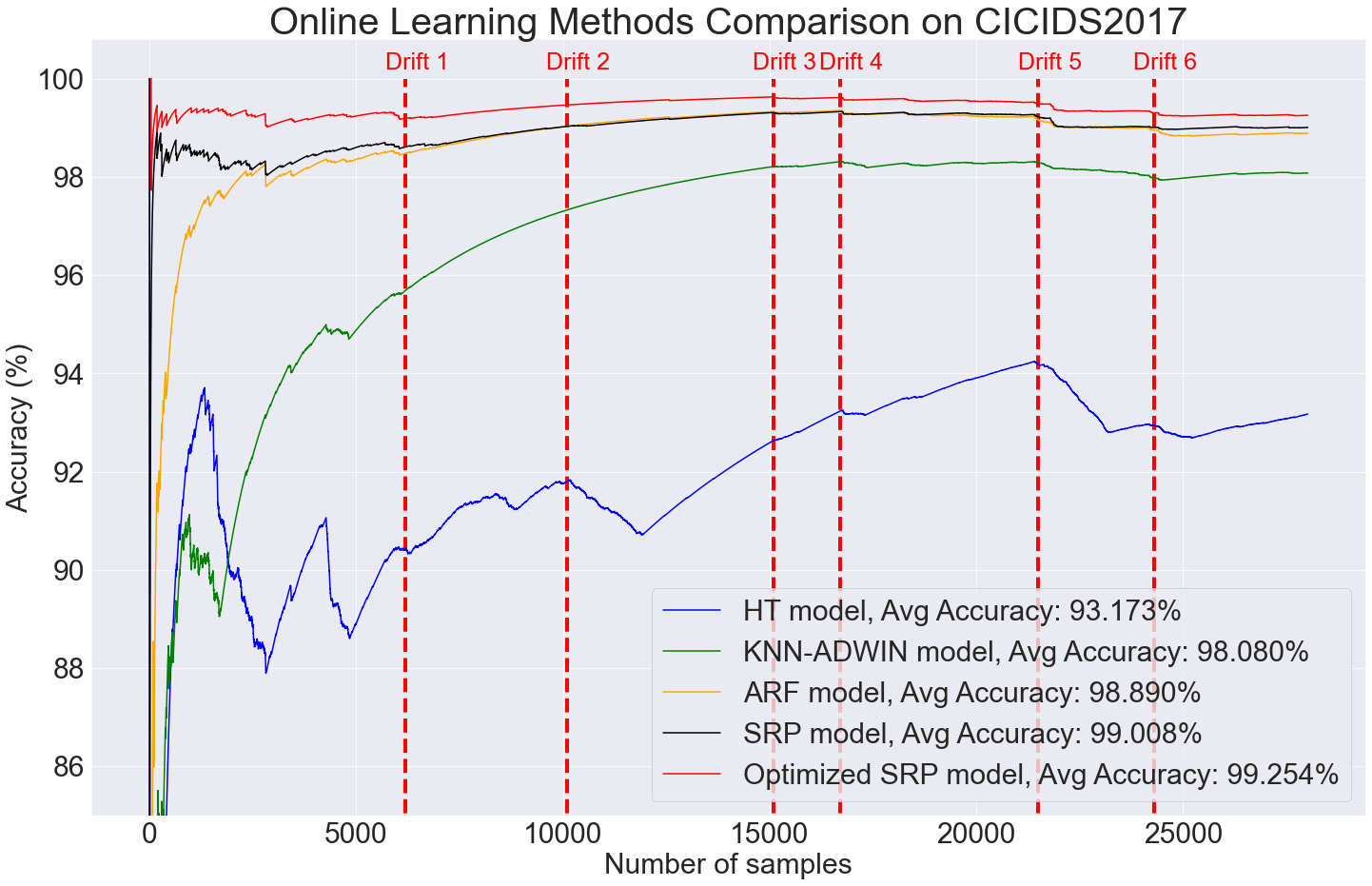}}
\caption{Performance comparison of online learning methods on the CICIDS2017 dataset.}
\label{cic}
\end{figure}

\begin{figure}[!t]
\centerline{
\includegraphics[width=\columnwidth]{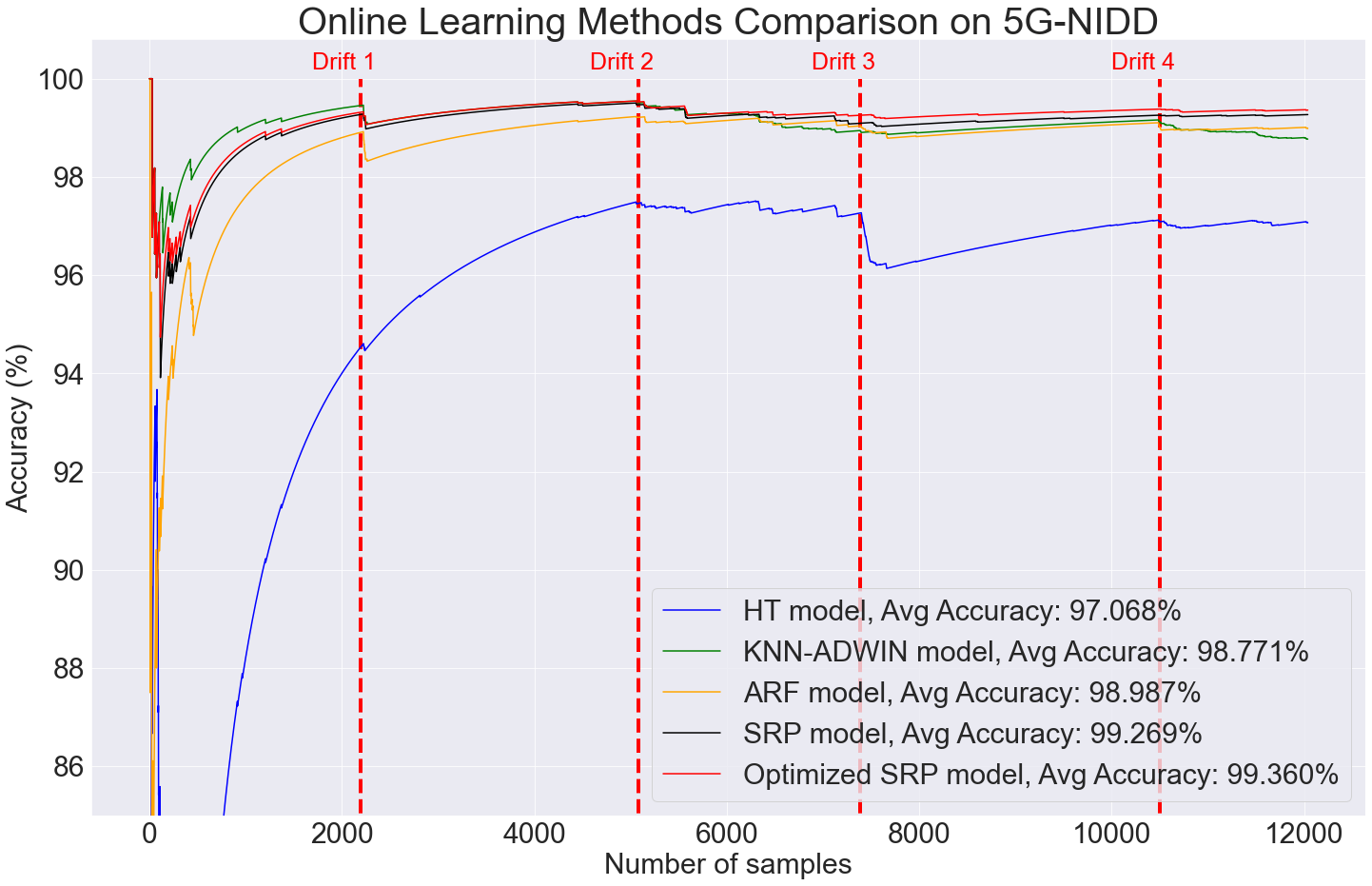}}
\caption{Performance comparison of online learning methods on the 5G-NIDD dataset.}
\label{5g}
\end{figure}

Secondly, to demonstrate the performance of the online AutoML-based IDS that incorporates automated model updating for dynamic ZTN environments, the experimental results are illustrated in Table \ref{result11} and Figs. \ref{cic} and \ref{5g}. The hyperparameters of the two better-performing online learning models, ARF and SRP, are optimized using PSO, as illustrated in Table \ref{hpo}.

As shown in Table \ref{result11} and Fig. \ref{cic}, all four base online learning models—HT \cite{drift3}, KNN-ADWIN \cite{knnadwin}, ARF \cite{arf}, and SRP \cite{srp}—can adapt to model drifts and restore their accuracy on the CICIDS2017 dataset. Among the base online learning models after implementing AutoDP and AutoFE, the ARF and SRP models achieve higher accuracy of 98.890\% and 99.008\% and show notable adaptability to model drifts, as indicated by the accuracy dips and recoveries at drift points in Figure \ref{cic}. The SRP model, particularly its optimized variant, demonstrates remarkable resilience, maintaining an average accuracy of 99.254\%. This is closely followed by the ARF model with an average accuracy of 99.090\%. These results illustrate the potential of AutoML-enhanced models in maintaining high detection accuracy amidst the data volatility characteristic of cybersecurity threats. 

Moreover, when compared to online learning models without the integration of AutoDP and AutoFE (where AutoML Procedures are not applied), the accuracy and F1-scores for all four models observed a modest improvement, attributable to the data quality enhancement and feature selection refinement facilitated by these procedures. Notably, the model learning times for these models saw a significant decrease, with the SRP model showing a remarkable reduction from 80.0s to 58.4s. This reduction was primarily driven by the efficiency of the automated feature selection process. Among these, the optimized SRP model stood out, achieving the highest accuracy of 99.254\% and an F1-score of 99.093\% on the CICIDS2017 dataset, while maintaining a reduced model learning time of 68.2s. The optimized models exhibit improved detection capabilities compared to the non-optimized models. This improvement is achieved through the AutoML pipeline, which fine-tunes model parameters and structures to better suit the data characteristics, as evidenced by the hyperparameter configurations detailed in Table \ref{hpo}. These configurations are the result of an extensive search within predefined ranges, guided by the AutoML process to identify the most effective values for each parameter.

The online learning results on the 5G-NIDD dataset results, as presented in Table \ref{result11} and illustrated in Fig. \ref{5g}, further underscore the efficacy of incorporating AutoML techniques, specifically AutoDP and AutoFE, in enhancing the performance of online learning models within dynamic ZTN environments. For the 5G-NIDD dataset, the application of AutoDP and AutoFE procedures yielded a notable improvement in both accuracy and F1-scores across all four online learning models—HT, KNN-ADWIN, ARF, and SRP. This enhancement can be attributed to the refined data quality and improved feature selection facilitated by these AutoML procedures. Among the models, the SRP model, with AutoML optimizations, exhibited a significant performance leap, achieving an exceptional average accuracy of 99.360\% and an F1-score of 99.473\%. This not only highlights the SRP model's superior adaptability to evolving network conditions but also its capability to maintain high detection accuracy amidst the complex data landscape of 5G networks.

The ARF model also showed commendable performance, with an improved average accuracy of 99.227\%, demonstrating its robustness in the face of AML attacks and its responsiveness to the dynamic nature of 5G network traffic. The observed reduction in model learning time, especially for the SRP model—from 45.4s to 24.0s in the presence of AutoDP and AutoFE—emphasizes the impact of automated feature selection in streamlining the model optimization process.

The superiority of the optimized SRP model on the CICIDS2017 and 5G-NIDD datasets, alongside its counterpart ARF model, reiterates the critical role of AutoML in tailoring IDS solutions to specific network environments. The enhanced detection capabilities of these models, as evidenced by their performance metrics, affirm the potential of AutoML to revolutionize IDS development by automating the tuning of model parameters and structures to optimally match the characteristics of the data being analyzed. This adaptability is essential for maintaining the efficacy of autonomous IDSs in the face of the constantly evolving threat landscape characteristic of cybersecurity challenges in ZTNs that have dynamic networking environments and high-level network automation requirements.

The optimized models can achieve high performance without overfitting mainly due to the following reasons \cite{mymth}: 
\begin{enumerate}
\item The proposed models were evaluated using 5-fold cross-validation to mitigate overfitting, providing a more accurate representation of their generalizability across unseen data. This method helps in ensuring that our high performance is consistent across various subsets of the data.
\item A comprehensive AutoFE method has been implemented to enhance the models' ability to generalize. By systematically removing irrelevant and misleading features, the proposed models are trained only on data that contribute to predictive performance, reducing the risk of overfitting.
\item The datasets used, namely CICIDS2017 and 5G-NIDD, have distinct attack patterns that are easier to distinguish compared to more complex datasets. This inherent characteristic, coupled with our robust model evaluation and feature engineering methods, contributes to the high accuracies reported. Comparable studies in the field have achieved similar performance, indicating that our results are in line with expectations for these datasets. For instance, as described in Section \ref{S3-sec}, the DL-LSTM model proposed in \cite{ids3} and the PyDSC-IDS model proposed in \cite{ids4} achieved high accuracy and F1-scores of more than 99.3\% on the CICIDS2017 dataset. Similarly, the Embeddings \& FC model, proposed in \cite{ids2}, showcased a remarkable accuracy of 99.123\% and an F1-score of 98.666\% on the 5G-NIDD dataset. This further highlights that the two datasets have distinct attack characteristics that can be easily identified, which directly impacts the high performance of ML models.

\end{enumerate}

\subsection{Use Case 2: AML Attacks and Defense} \label{S6-U2}

\subsubsection{Experimental Setup} \label{S6-U2-pro}

Although the autonomous IDSs developed in the first case study achieved high performance for ZTNs, the network automation requirements of ZTNs have introduced new security threats, AML attacks, due to the lack of human supervision. The second case study implements a cyber-defense exercise against three common AML attacks discussed in Section \ref{S5-evasion} (\textit{i.e.}, DTA \cite{dta}, FGSM \cite{fgsm}, and BIM \cite{bim}). These three AML attacks are launched in separate experiments to generate adversarial samples. These adversarial samples are intended to disrupt the input data used by the AutoML-based IDS obtained in the first case study. The purpose of these experiments is to demonstrate the potential damage caused by AML attacks on ML models within ZTNs/6G networks.

Subsequently, the experiments progress to the next stage, where a fundamental AML defense mechanism, namely adversarial sample detection and filtering \cite{zsmsec3}, is deployed against each type of AML attack. This illustrates the necessity and effectiveness of this fundamental defense mechanism in mitigating AML attacks and maintaining the reliability of the AutoML-based IDS within ZTNs. In this case study, the optimized LightGBM models fulfill dual roles as both the models are susceptible to compromise under AML attacks and the defense models for detecting adversarial samples.

The experiments in the second case study were conducted on the same hardware as the first, utilizing a machine with an i7-8700 processor and 16 GB of RAM. This machine represents a ZTN server equipped with autonomous IDSs, now targeted by AML attacks intended to disrupt IDS functionality. The AML attacks and defense mechanisms were implemented by extending Python libraries: LightGBM \cite{lightgbm} for the compromised ML model and adversarial sample detection model development, and the Adversarial Robustness Toolbox (ART) \cite{art} for simulating three adversarial attacks: DTA \cite{dta}, FGSM \cite{fgsm}, and BIM \cite{bim}.

The specific procedures of the proposed cyber-defense exercise are as follows:
\begin{enumerate}[label=\roman*)]
\item Evaluate the optimized IDS model obtained from the first case study's AutoML procedure, LightGBM with the optimal hyperparameters outlined in Table \ref{hpo}, on the original data set. This serves as the baseline model without any AML attacks. \label{exercise-1}
\item Generate adversarial samples using one of the AML attacks (\textit{i.e.}, DTA, FGSM, or BIM) and combine these samples with the original training set. This simulates a disrupted input dataset under the AML attack. \label{exercise-2}
\item Assess the IDS model on the adversarial samples generated in Procedure \ref{exercise-2}. This demonstrates the performance of a model under an AML attack. \label{exercise-3}
\item Develop an adversarial sample detection model using the same LightGBM model. This model is trained on an integrated dataset of the original training set and adversarial samples to distinguish between them and identify adversarial samples. \label{exercise-4}
\item Remove the detected adversarial samples from the training set to create a sanitized dataset. \label{exercise-5}
\item Re-train the IDS model on the sanitized training set and evaluate its performance on the test set, representing the effectiveness of the chosen defense model: adversarial sample detection and filtering. \label{exercise-6}
\end{enumerate}

\subsubsection{Experimental Results and Analysis}

\begin{table*}[htbp]
\setlength\extrarowheight{1pt}
\centering%
\caption{The experimental results of the AML attack and defense models on the CICIDS2017 and 5G-NIDD datasets.}
\scalebox{0.85}{
\begin{tabular}{|>{\centering\arraybackslash}p{1.2cm}|>{\centering\arraybackslash}p{1.2cm}|>{\centering\arraybackslash}p{3.0cm}|>{\centering\arraybackslash}p{1.1cm}|>{\centering\arraybackslash}p{1cm}|>{\centering\arraybackslash}p{0.8cm}|>{\centering\arraybackslash}p{0.8cm}|>{\centering\arraybackslash}p{1.2cm}|>{\centering\arraybackslash}p{1.1cm}|>{\centering\arraybackslash}p{1cm}|>{\centering\arraybackslash}p{0.7cm}|>{\centering\arraybackslash}p{0.7cm}|>{\centering\arraybackslash}p{1.2cm}|}

\hline
\multirow{2}{*}{\textbf{Model}} & \multirow{2}{*}{\textbf{State}} & \multirow{2}{*}{\textbf{Model Specifics}} & \multicolumn{5}{c|}{\textbf{CICIDS2017 Dataset}} & \multicolumn{5}{c|}{\textbf{5G-NIDD Dataset}} \\ \cline{4-13} 
                                &                                 &                                           & \textbf{Accuracy (\%)} & \textbf{Precision (\%)} & \textbf{Recall (\%)} & \textbf{F1 (\%)} & \textbf{Model Learning Time (s)} & \textbf{Accuracy (\%)} & \textbf{Precision (\%)} & \textbf{Recall (\%)} & \textbf{F1 (\%)} & \textbf{Model Learning Time (s)} \\ \hline
\multirow{10}{*}{\shortstack{The\\\\ AutoML-\\\\ based IDS \\\\Model}} & Original                        & The original   IDS model                                  & {99.823}                 & {99.468}                  & {99.644}               & {99.556}           & 0.2                              & {99.877}                 & {99.799}                  & {100.0}                & {99.899}           & 0.7                              \\ \cline{2-13} 
                                              & \multirow{6}{*}{\shortstack{Under\\\\ Attack}}   & The IDS model   under the DTA attack                      & {78.608}                 & {15.873}                  & {1.778}                & {3.197}            & 0.1                              & {62.459}                 & {68.947}                  & {70.383}               & {69.658}           & 0.6                              \\ \cline{3-13} 
                                              &                                 & The IDS model   under the FGSM attack                     & {86.310}                 & {96.296}                  & {32.356}               & {48.436}           & 0.3                              & {40.584}                 & {97.826}                  & {3.022}                & {5.863}            & 0.5                              \\ \cline{3-13} 
                                              &                                 & The IDS model   under the BIM attack                      & {88.324}                 & {96.586}                  & {42.756}               & {59.273}           & 0.5                              & {38.734}                 & {33.333}                  & {0.067}                & {0.134}            & 0.7                              \\ \cline{2-13} 
                                              & \multirow{6}{*}{Recovered}      & The IDS model recovered   from the DTA attack             & {99.806}                 & {99.379}                  & {99.644}               & {99.512}           & 0.2                              & {99.877}                 & {99.799}                  & {100.0}                & {99.899}           & 0.5                              \\ \cline{3-13} 
                                              &                                 & The IDS model recovered   from the FGSM attack            & {99.770}                 & {99.291}                  & {99.556}               & {99.423}           & 0.2                              & {99.877}                 & {99.799}                  & {100.0}               & {99.899}           & 0.8                              \\ \cline{3-13} 
                                              &                                 & The IDS model recovered   from the BIM attack             & {99.806}                 & {99.379}                  & {99.644}               & {99.512}           & 0.3                              & {99.836}                 & {99.732}                  & {100.0}                & {99.866}           & 0.6                              \\ \hline
\multirow{7}{*}{\shortstack{The AML \\\\Detection\\\\ Model}}    & \multirow{7}{*}{-}              & The adversarial   sample detection model for DTA attacks  & {99.937}                 & {99.932}                  & {99.942}               & {99.937}           & 0.3                              & {99.822}                 & {99.720}                  & {99.925}               & {99.822}           & 0.6                              \\ \cline{3-13} 
                                              &                                 & The adversarial   sample detection model for FGSM attacks & {99.407}                 & {99.667}                  & {99.144}               & {99.405}           & 0.3                              & {99.663}                 & {99.775}                  & {99.550}               & {99.662}           & 0.5                              \\ \cline{3-13} 
                                              &                                 & The adversarial   sample detection model for BIM attacks  & {99.453}                 & {99.673}                  & {99.230}               & {99.451}           & 0.4                              & {99.634}                 & {99.579}                  & {99.691}               & {99.635}           & 0.5                              \\ \hline
\end{tabular}
}
\label{result2}
\end{table*}

\begin{figure*}[!t]
\centerline{
\includegraphics[width=16.5cm]{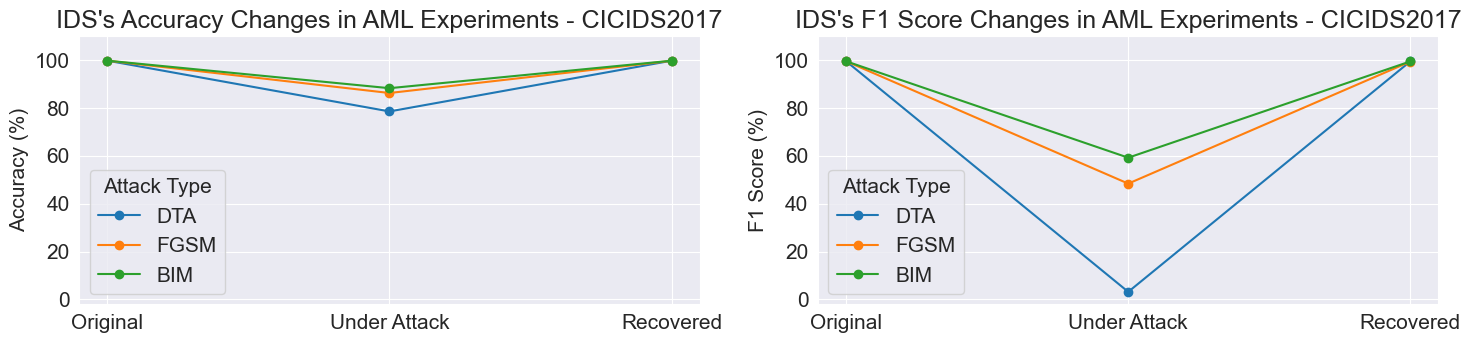}}
\caption{The performance changes of the IDS model on the CICIDS2017 dataset in the AML attacks and defense experiments.}
\label{amlcic}
\end{figure*}

\begin{figure*}[!t]
\centerline{
\includegraphics[width=16.5cm]{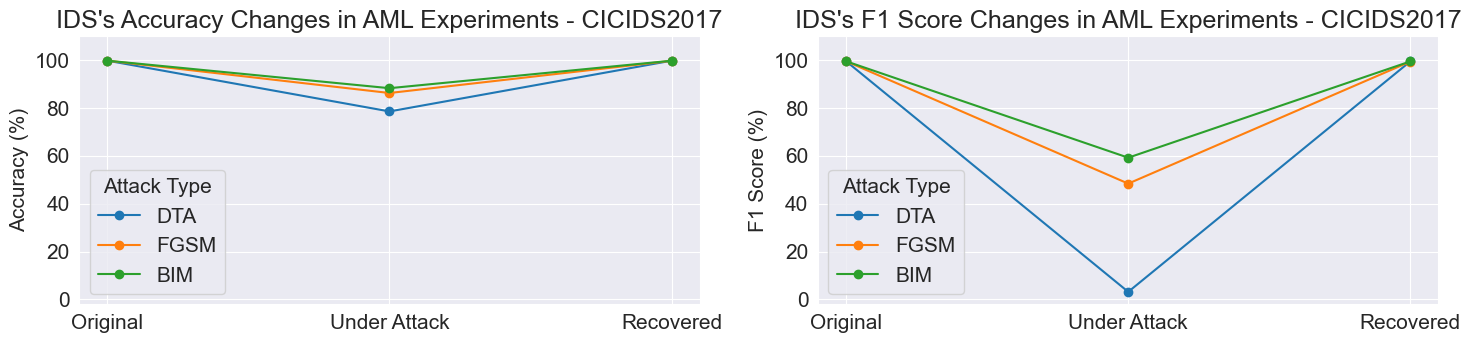}}
\caption{The performance changes of the IDS model on the 5G-NIDD dataset in the AML attacks and defense experiments.}
\label{aml5g}
\end{figure*}

The performance of the cyber-defense exercise against AML attacks was evaluated by comparing the accuracy, precision, recall, F1-score, and model learning time of both the AutoML-based IDS model and the AML detection model, using the CICIDS2017 and 5G-NIDD datasets. The results for these two datasets are presented in Table \ref{result2}. Figures \ref{amlcic} and \ref{aml5g} demonstrate the accuracy and F1-score changes of the IDS model on the two datasets during the AML attacks and after recovery. 

Table \ref{result2} illustrates the results of two model groups: the AutoML-based IDS models and the AML detection models, on the two datasets. The results of the AutoML-based IDS models can be classified into three states: the original IDS models, the IDS models under attack, and the recovered IDS models. These states align with the proposed cyber-defense exercise procedures \ref{exercise-1}, \ref{exercise-3}, and \ref{exercise-6}, discussed in Section \ref{S6-U2-pro}.

The results of the first state present the performance of the original IDS model as a baseline. The results of the second state display the performance of the IDS model under different AML attacks, showcasing the impact or damage of these attacks on the IDS model's performance. The third state’s results demonstrate the performance of the IDS model after implementing the adversarial sample detection and filtering mechanism, illustrating the effectiveness of this defense mechanism in restoring the IDS model's performance after an AML attack. The final part of the results illustrates the performance of the AML attack detection models, aligning with the adversarial sample detection model derived from the proposed cyber-defense exercise procedure \ref{exercise-4} from Section \ref{S6-U2-pro}.

The cyber-defense exercise begins with the evaluation of the original IDS model, followed by the IDS model performance assessment under DTA, FGSM, and BIM attacks.  Subsequently, the adversarial sample detection and filtering mechanism is applied to remove detected adversarial samples, and the IDS model is re-trained and evaluated on the sanitized dataset.
Table \ref{result2} encapsulates the results of the cyber-defense exercise using the CICIDS2017 and 5G-NIDD datasets. For the original IDS model without AML attacks, the model demonstrates a high degree of performance across all metrics: accuracy, precision, recall, and F1-score. Specifically, the F1-scores of the pristine IDS model stand at 99.556\% and 99.899\% on the CICIDS2017 and 5G-NIDD datasets, respectively. 

However, the performance of the IDS model significantly decreases under the influence of DTA, FGSM, and BIM attacks on both datasets. The F1-scores of the IDS fall to a range between 0.134\% and 69.658\%, reflecting a decline of at least 30\% across both datasets. As shown in Table \ref{result2}, among the three AML attacks, DTA causes the most damage to the CICIDS2017 dataset, while BIM proves to be the most effective AML attack against the 5G-NIDD datasets. In addition, BIM attacks cause a significant increase in the model learning time for both datasets.

In contrast, the implementation of the adversarial sample detection and filtering mechanism results in a remarkable performance improvement for the IDS model. The accuracy, precision, recall, and F1-scores exhibit significant enhancements, closely resembling those of the original IDS model. On the CICIDS2017 dataset, the F1-scores of the restored IDS models range from 99.423\% to 99.512\%, whereas on the 5G-NIDD dataset, they range from 99.866\% to 99.899\%.

The performance of the AML detection model, as depicted in Table \ref{result2}, demonstrates its effectiveness in identifying and differentiating adversarial samples from the original, legitimate samples. For each form of AML attack, high accuracy, precision, recall, and F1-scores are recorded, ranging from 99.451\% to 99.937\%.

In summary, the destructiveness of AML attacks on ML models within ZTNs can be seen in the following:
\begin{enumerate}[label=\roman*)]
\item \textit{Reduced Accuracy and Effectiveness}: AML attacks can significantly degrade the accuracy of ML-based IDSs, rendering them ineffective at detecting and mitigating security threats. This poses a threat to network resources and infrastructure.
\item \textit{Increased Complexity}: AML attacks add an additional layer of complexity to the already complex security landscape of ZTNs, necessitating the development of defense mechanisms to defend against these sophisticated threats.
\item \textit{Erosion of Trust}: Successful AML attacks can erode confidence in ML models and their capacity to provide reliable security solutions for ZTNs, resulting in decreased confidence in the adoption of these technologies.
\end{enumerate}

Additionally, the adversarial sample detection and filtering method utilized in the case study can effectively distinguish adversarial samples generated by AML attacks and remove them. This results in a clean training set capable of effectively restoring the performance of ML models developed in the first case study.

\section{Open Challenges and Research Directions} \label{S7}
Despite the existence of various network security solutions, effectively deploying them within ZTNs poses several unresolved challenges. In this Section, the open challenges and research directions concerning ZTN security mechanisms are discussed. These challenges and directions are comprehensively summarized in Table \ref{7-challenges}.

\begin{table*}[htbp]
\setlength\extrarowheight{1pt}
\centering
\caption{The open challenges and research directions of ZTNs/6G network security.}
\begin{tabular}{|>{\centering\arraybackslash}p{1.5cm}|>{\centering\arraybackslash}p{3cm}|p{11.8cm}|}
\hline
\textbf{Category} & \textbf{Challenge} & \multicolumn{1}{>{\centering\arraybackslash}p{11.8cm}|}{\textbf{Brief Summary}} \\
\hline
\multirow{7}{*}{\shortstack{General\\\\ Cybersecurity\\\\ Challenges}} & Rapid Evolvement of Cyber-Attacks \cite{zsmsec3}  & The constantly evolving nature of cyber threats and the use of AI/ML by cybercriminals demand continuous updating and deployment of new security measures. \\
\cline{2-3}
& Network Complexity \cite{aml1} & The complexity of 5G and 6G networks, with numerous connected devices and systems, makes threat detection and response increasingly difficult. \\
\cline{2-3}
& Cybersecurity System Cost \cite{cost}& Implementing effective security measures can strain resources, particularly for small enterprises. \\
\cline{2-3}
& Cross-Layer Network Security \cite{zsm4}& Cross-layer intelligence in ML-based network management frameworks enhances overall network security and resilience by leveraging interdependencies and cooperation across protocol stack layers. \\
\hline
\multirow{7}{*}{\shortstack{AML\\\\ Defense\\\\ Challenges}} & Lack of Autonomous Solutions \cite{zsm1} & Developing autonomous cybersecurity solutions is crucial for 6G networks, but current automation techniques require additional research to address AI/ML model security challenges. \\
\cline{2-3}
& Feasible Cyber-Defense Exercises \cite{dt1} & Digital Twin (DT) \cite{dt2} technology can facilitate cyber-defense exercises without disrupting network operations, enabling the evaluation and improvement of AML protection models. \\
\cline{2-3}
& Model Privacy Issues \cite{6gpri1} & Sharing data and models between network devices raises privacy concerns; federated learning \cite{fl} and blockchain \cite{blockchain} may address these issues, but require further research. \\
\cline{2-3}
& Interpretability and Explainability \cite{xai} & Incorporating Explainable AI (XAI) techniques into AML protection models can improve understanding of decision-making processes, enabling more robust defenses against adversarial attacks. \\
\hline

\multirow{5}{*}{\shortstack{AutoML\\\\ Challenges}} &  Data Pre-Processing and Feature Engineering \cite{automl7} & Automating data pre-processing and feature engineering in AutoML pipelines is complex, requiring additional research. \\
\cline{2-3}
& Large Scale AutoML \cite{myhpo} & Applying AutoML to large-scale data remains an open challenge, as most AutoML solutions are designed for small datasets. \\
\cline{2-3}
& Trustworthy AI \cite{trust1}-\cite{trust2} & Developing AI systems that are reliable, safe, secure, and ethical is crucial for AutoML models, and requires addressing aspects like robustness, reproducibility, explainability, privacy, accountability, and fairness. \\
\hline
\end{tabular}
\label{7-challenges}
\end{table*}

\subsection{General Cybersecurity Mechanism Development Challenges} 

\subsubsection{Rapid Evolvement of Cyber-Attacks} 
One of the major challenges of general cybersecurity is the rapidly evolving nature of cyber threats. Cybercriminals are constantly developing new methods to bypass existing security measures, and organizations must be able to adapt to these changes to maintain their security. Moreover, attackers can leverage AI/ML techniques to prioritize vulnerabilities for massive network attacks. For instance, AI-based botnets can identify zero-day vulnerabilities in IoT devices and exploit them to initiate DDoS attacks against 5G RAN resources \cite{zsmsec3}.

This necessitates organizations to invest in the development and deployment of new cybersecurity technologies and to update existing security measures on a regular basis.

\subsubsection{Network Complexity} 
Another challenge of cybersecurity is the complexity of 5G and 6G networks \cite{aml1}. As the number of connected devices increases and cloud computing becomes prevalent, detecting and responding to cybersecurity threats in complex networks becomes increasingly difficult. Numerous systems and devices require security measures, and identifying the most vulnerable components can be challenging.
\subsubsection{Cybersecurity System Cost} 
The financial burden of cybersecurity poses a significant challenge, particularly for small enterprises. The cost of implementing effective security measures must be measured against the cost of a security violation \cite{cost}. Additionally, quantifying the cost of network security is challenging as it encompasses not only the expenses related to implementing security measures but also potential costs arising from data breaches, delays, and reputational damage.
\subsubsection{Cross-Layer Network Security} 
Traditional ML-based network management services focus on single protocol stack layers, limiting large-scale intelligence potential. Cross-layer cooperation, however, enables the development of more flexible and efficient solutions in heterogeneous networks \cite{zsm4}. In the context of cross-layer network security, it is essential to examine the interdependencies between different layers of the protocol stack. For instance, the RAN can optimize wireless channel operation by dynamically adapting both the link and physical layers accordingly. By leveraging cross-layer intelligence, it becomes possible to identify and address security challenges that span multiple layers, leading to more comprehensive and robust security mechanisms.

Cross-layer network security solutions can also facilitate more effective detection and response to cybersecurity threats, as the integration of information from different layers can provide a more accurate and complete understanding of the network's security state. This holistic approach can lead to the development of more advanced ML-based security models that can adapt to evolving threats, improving the overall resilience of the network.

\subsection{AML Defense Model Development Challenges} 

\subsubsection{Lack of Autonomous Solutions} 
Existing AML security models typically require human analysis or supervision to comprehend attack patterns and design effective countermeasures. However, autonomous cybersecurity solutions are required to facilitate network automation for ZTNs/6G networks \cite{zsm1}. AutoML techniques discussed in Section \ref{S5-AutoML} are promising solutions for interacting with networking environments and developing security models automatically. However, current AutoML techniques are proposed for general cybersecurity issues, and additional research is necessary to address AI/ML model security challenges.   
\subsubsection{Feasible Cyber-Defense Exercises} 
Cyber-defense exercises that simulate cyber-attacks in networks are often required to develop and evaluate AML countermeasures and security mechanisms, but launching attacks will disrupt the normal operation of physical networks \cite{dt1}. Digital Twin (DT) \cite{dt2}  is a promising technology for improving traditional cyber-defense exercises. The primary objective of DT models is to generate virtual copies of a physical system or network to simulate its behavior and real-time usage. DT models enable cyber-defense exercises and security model evaluations in cybersecurity applications without interfering with the normal operation of networks. In DT-assisted networks,  AML attacks and defense operations can be simulated by launching multiple types of AML attacks and then evaluating the effectiveness of protection mechanisms. Once their performance is evaluated, the defensive models will be transmitted back to the actual network from the virtual networks.
\subsubsection{Model Privacy Issues} 
The implementation of certain AML countermeasures for 5G/6G networks raises concerns regarding model privacy, as it necessitates the sharing of data and models from multiple network devices to develop comprehensive security measures \cite{6gpri1}. Developing effective countermeasures for specific attacks usually requires an exhaustive and detailed comprehension of AML attacks and affected ML models. However, data or model leakage may occur during the development of AML security models. Federated learning \cite{fl} and blockchain \cite{blockchain} techniques are potential solutions to address the privacy issues of AML security model development, but their integration with AML security approaches requires additional research. 
\subsubsection{Interpretability and Explainability}
Explainable Artificial Intelligence (XAI)  is a crucial aspect of developing and deploying AML protection models in ZTNs \cite{xai}. Even though ML models have the potential to achieve high levels of accuracy and performance, their decision-making processes are often viewed as "black boxes", which makes it challenging for users to understand the reasoning behind their conclusions. The limited interpretability of AML models can impede the development of effective countermeasures, as security analysts face challenges in identifying potentially exploitable vulnerabilities within the model. Incorporating XAI techniques into AML protection models can provide a deeper understanding of the decision-making process, enabling the identification of vulnerabilities and the development of more robust defenses against adversarial attacks. Common XAI methods include feature importance calculation \& analysis, visualization methods, and model agnostic techniques, such as SHapley Additive exPlanations (SHAP) and Local Interpretable Model-agnostic Explanations (LIME) \cite{xai2}. However, achieving the optimal balance between model performance, interpretability, and security remains a challenging task that necessitates further research. Future work in XAI is essential to enhance the security of AI and ZTNs against emerging cyber threats, ensuring they are reliable and safe for supporting advanced functionalities services.

\subsection{AutoML Technology Challenges} 
During the past decade, AutoML has made substantial progress in automating model construction and development, especially for supervised learning tasks. However, AutoML faces several challenges that need to be addressed before it can be widely applied to real-world ZTN/6G network applications \cite{automl7}:
\subsubsection{Automated Data Pre-Processing and Feature Engineering} 
Although there are many existing AutoML solutions, the vast majority of them concentrate on automated model selection and HPO, with less emphasis placed on automated data pre-processing and feature engineering \cite{automl7}. Yet, these two aspects are vital components of the AutoML pipeline and directly influence the performance of systems, particularly in the context of B5G/6G networks that require large-scale data analytics. Feature engineering is especially challenging to automate and generalize, as it heavily depends on the specific task and dataset \cite{automl5}. Implementing effective feature engineering often necessitates specialized domain knowledge or considerable effort. As such, automated feature engineering, despite its complexity, is an imperative research topic that warrants further investigation.
\subsubsection{Large Scale AutoML} 
AutoML's application to large-scale data remains unresolved. Since AutoML pipelines often need a large number of model trainings to determine the optimal end learner, most AutoML solutions are created on small datasets, with just a handful capable of large-scale data learning. For example, due to the huge size of the ImageNet dataset, research on AutoML solutions for the ImageNet challenge is still limited \cite{myhpo}. 
\subsubsection{Trustworthy AI} 
Trustworthiness of AI/ML techniques refers to the extent to which an AI system is reliable, safe, secure, and ethical \cite{trust1}. The development of trustworthy AI systems is crucial for ensuring that AI can be used for the benefit of ZTNs rather than causing harm. Trustworthy AI is particularly important for AutoML models, as they often lack direct human supervision. Key aspects of AI trustworthiness include robustness, reproducibility, explainability, privacy, accountability, and fairness \cite{trust2}.
\begin{enumerate}[label=\roman*)]
\item \textit{Robustness}: It is a key aspect of AI systems, which refers to their ability to handle execution errors, incorrect inputs, and unseen data. A robust AI system should exhibit resilience at different levels of data, algorithms, and systems. 
\item \textit{Reproducibility}: It indicates the ability of AI/ML models to be reproduced. It is crucial for validating and assessing AI/ML research, enabling the identification and mitigation of potential risks in AI systems. The AI research community increasingly views reproducibility as a requirement for publishing research.
\item \textit{Explainability}: It is essential for building trust in AI/ML technology and entails understanding the decision-making process of AI/ML models \cite{trust3}. Research on AI/ML explainability has been conducted in two aspects. The first aspect involves the development of fully or partially explainable ML models, such as linear regression and decision tree-based models. The second aspect focuses on analyzing complex models like DL models by examining their input, intermediate results, and output. As described in Section VII-B.4, XAI is a critical research topic to enhance the trustworthiness of AutoML and ZTNs. 
\item \textit{Privacy Protection}: It refers to safeguarding data that can directly or indirectly identify individuals or households. Privacy protection spans the entire lifecycle of AI/ML systems, covering data collection \& pre-processing, model training, and deployment.
\item \textit{Accountability}: It is essential for ensuring that AI systems adhere to trustworthiness requirements. Accountability incorporates the entire lifecycle of an AI system and requires stakeholders to justify their design, implementation, and operation in accordance with human values.
\item \textit{AI Fairness}: It involves mitigating various types of bias, such as data bias, model bias, and procedural bias \cite{trust4}. These biases frequently lead to the unjust treatment of various groups based on their protected information, such as gender, race, and ethnicity. To prevent the perpetuation or exacerbation of social bias, it is vital to resolve impartiality in AI systems.
\end{enumerate}

In conclusion, establishing trust in AI/ML and AutoML technologies is a critical step in their successful application to ZTNs. 

\section{Conclusion} \label{S8}
The growing demand for network automation in the next generation of networks has led to the development of Zero-Touch Networks (ZTNs), where Artificial Intelligence (AI) and Machine Learning (ML) play a critical role. However, successful deployment of ZTNs requires resolving numerous security challenges during their implementation. In this survey paper, we have conducted a comprehensive review of the security issues and vulnerabilities associated with ZTNs and explored potential solutions, with a specific focus on Automated ML (AutoML) technologies and Adversarial ML (AML) attacks \& defense mechanisms. Two case studies are presented in this paper. The first case study focuses on the development of autonomous frameworks to address security issues in ZTNs, demonstrating the effectiveness of AutoML technologies in securing ZTNs. The second case study on the cyber-defense exercises of AML attacks has been discussed to illustrate the devastating impact of AML attacks, along with potential defense mechanisms. Moreover, open challenges and future research directions for ZTN security research have been discussed, highlighting the need for ongoing research and innovation in this field.


\begin{IEEEbiography}[{\includegraphics[width=1in,height=1.25in,clip,keepaspectratio]{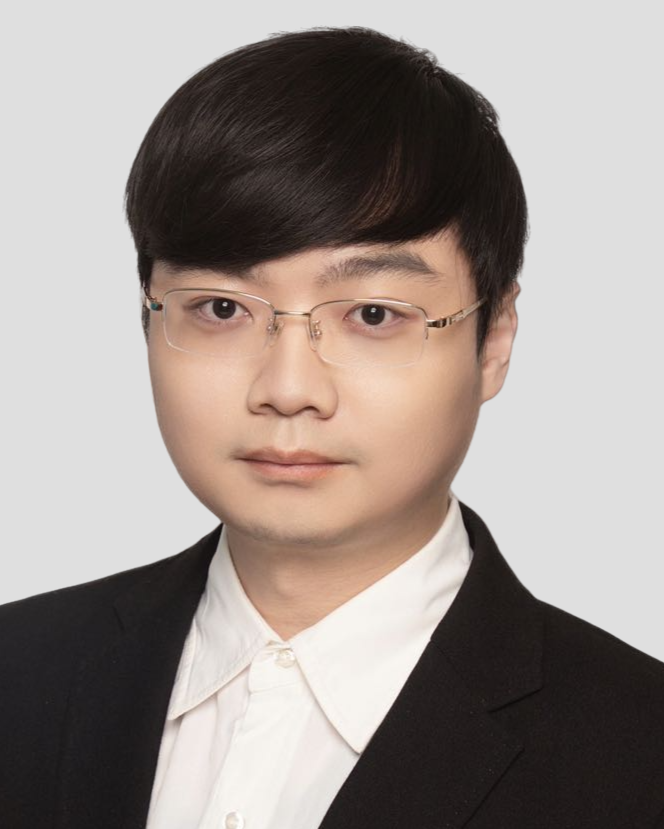}}]{Li Yang}\,is currently an Assistant Professor in the Faculty of Business and Information Technology at Ontario Tech University, and an Adjunct Research Professor in the Department of Electrical and Computer Engineering at Western University. He received his Ph.D. in Electrical and Computer Engineering from Western University in 2022. He was the vice chair of the IEEE Computer Society, London Section, Canada, from 2022 to 2023. He was also on the technical program committee for IEEE GlobeCom 2023 and 2024, the workshop chair for SMC-IoT 2023, and the technical session chair for IEEE CCECE 2020. His paper and code publications have received thousands of Google Scholar citations and GitHub stars. His research interests include cybersecurity, machine learning, deep learning, AutoML, model optimization, network data analytics, Internet of Things (IoT), intrusion detection, anomaly detection, concept drift, continual learning, and adversarial machine learning. Li Yang is also included in Stanford University/Elsevier's List of the World's Top 2\% Scientists. He was ranked among the world's Top 0.5\% of researchers in 'Networking \& Telecommunications' in 2024, and 52nd in Canada.
\end{IEEEbiography}

\begin{IEEEbiography}[{\includegraphics[width=1in,height=1.25in,clip,keepaspectratio]{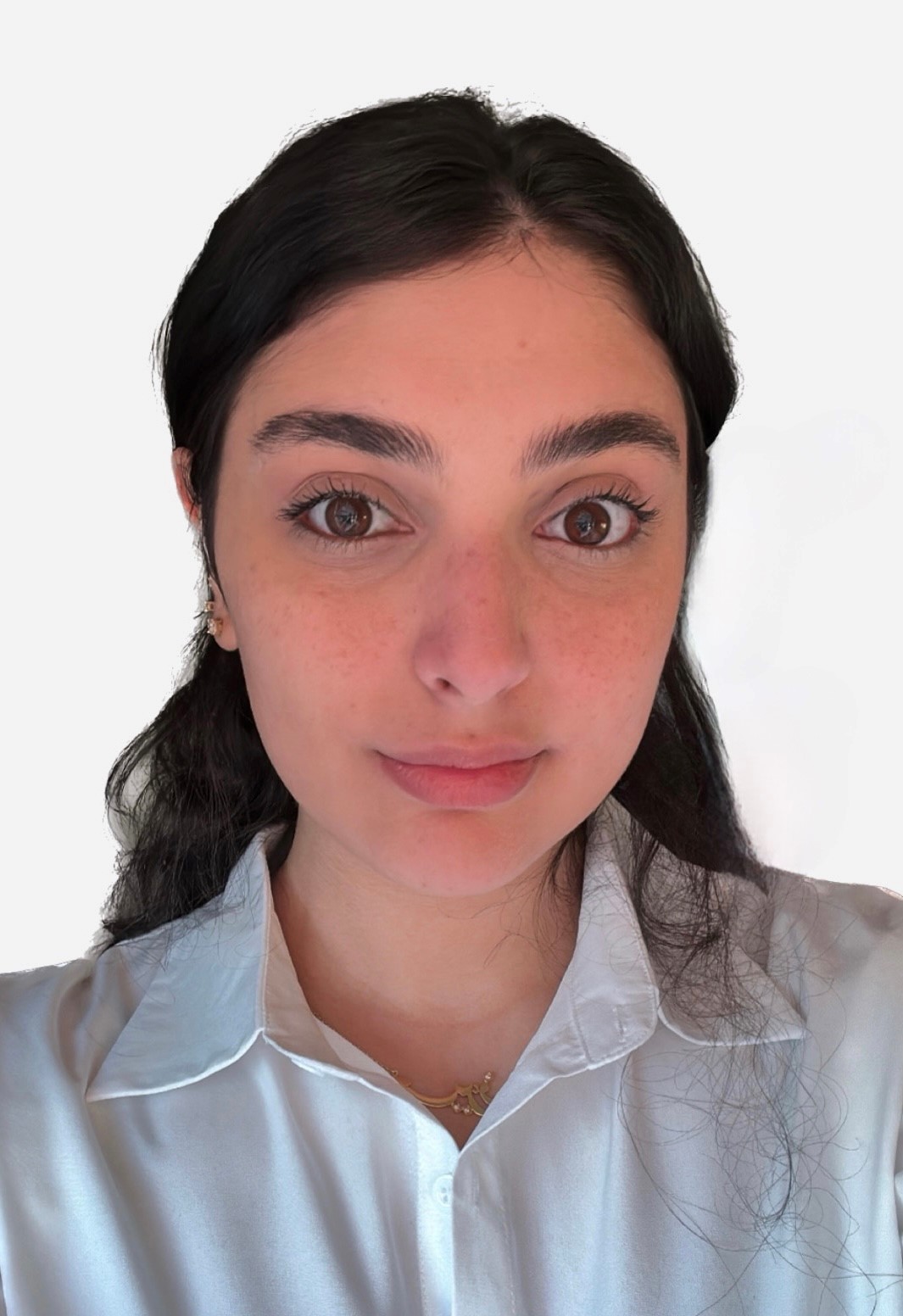}}]{Mirna El Rajab}\,received her MESc. in Electrical and Computer Engineering from Western University, London, ON, Canada in 2023. She completed her B.E. in Computer Engineering from Lebanese American University (LAU), Byblos, Lebanon, in 2021. She was a research assistant at the Department of Electrical and Computer Engineering, LAU, from May 2018 to April 2020. During that time, her work revolved around buffer-aided communication systems. Currently, her research interests lie in machine learning, zero-touch networks, and next-generation networks.
\end{IEEEbiography}

\begin{IEEEbiography}[{\includegraphics[width=1in,height=1.25in,clip,keepaspectratio]{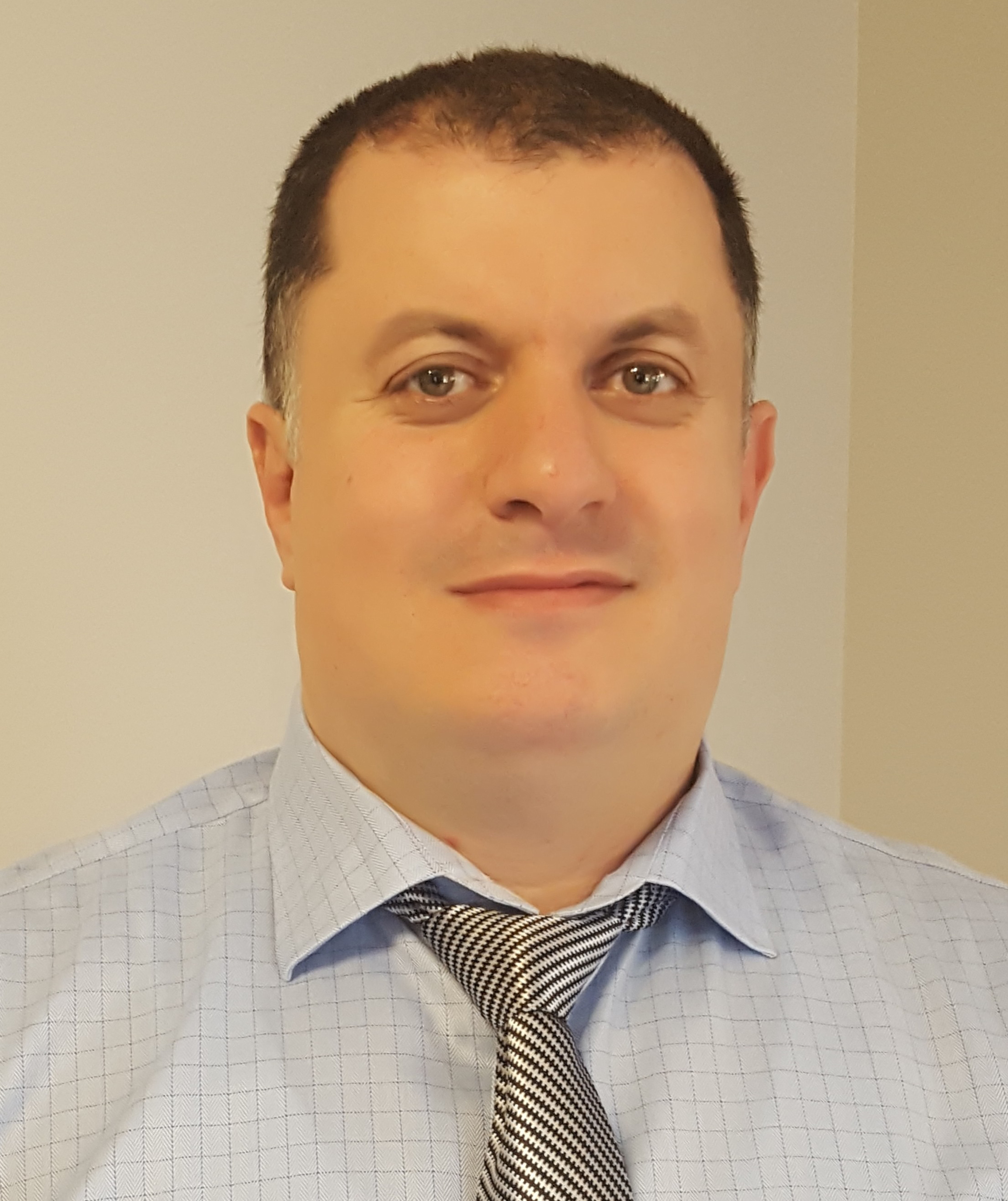}}] {Abdallah Shami}\,is currently a Professor and Chair of the Department of Electrical and Computer Engineering, Western University, London, ON, Canada, where he is also the Director of the Optimized Computing and Communications Laboratory. Dr. Shami has chaired key symposia for the IEEE GLOBECOM, IEEE International Conference on Communications, and IEEE International Conference on Computing, Networking and Communications. He was the elected Chair for the IEEE Communications Society Technical Committee on Communications Software and the IEEE London Ontario Section Chair. He is currently an Associate Editor of the IEEE Transactions on Information Forensics and Security, IEEE Transactions on Network and Service Management, and IEEE Communications Surveys and Tutorials journals. Dr. Shami is a Fellow of IEEE, a Fellow of the Canadian Academy of Engineering (CAE), and a Fellow of the Engineering Institute of Canada (EIC). 
\end{IEEEbiography}

\begin{IEEEbiography}[{\includegraphics[width=1in,height=1.25in,clip,keepaspectratio]{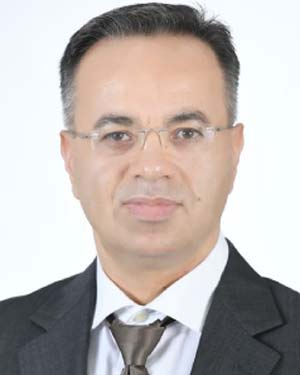}}] {Sami Muhaidat}\,received his Ph.D. in Electrical and Computer Engineering from the University of Waterloo, Ontario, in 2006. From 2007 to 2008, he was an NSERC Postdoctoral Fellow in the Department of Electrical and Computer Engineering at the University of Toronto, Canada. From 2008 to 2012, he served as an Assistant Professor in the School of Engineering Science at Simon Fraser University, British Columbia, Canada. Currently, he is a Professor and the Associate Dean for Research in the College of Computing and Mathematical Sciences at Khalifa University. He is also an Adjunct Professor at Carleton University, Ontario, Canada. Sami’s research interests include advanced digital signal processing techniques for wireless communications, intelligent surfaces, machine learning for communications, optical communications, and multiple-access techniques. He has served in various editorial roles, including as Area Editor for the IEEE Transactions on Communications, Guest Editor for the IEEE Network special issue on "Native Artificial Intelligence in Integrated Terrestrial and Non-Terrestrial Networks in 6G," and Guest Editor for the IEEE Open Journal of Vehicular Technology (OJVT) special issue on "Recent Advances in Security and Privacy for 6G Networks." Additionally, he has held positions as Senior Editor and Editor for IEEE Communications Letters, Editor for the IEEE Transactions on Communications, and Associate Editor for the IEEE Transactions on Vehicular Technology.
\end{IEEEbiography}

\vfill\pagebreak

\end{document}